\documentclass[%
 reprint,
superscriptaddress,
 amsmath,amssymb,
 aps,
pra,
]{revtex4-2}
\usepackage{xcolor}
\usepackage{graphicx}
\usepackage{dcolumn}
\usepackage{bm}
\usepackage{bbm}
\usepackage{relsize}
\usepackage{romannum}
\usepackage{csquotes}
\usepackage{xcolor}
\usepackage{soul}
\usepackage{floatrow}
\usepackage{tikz}
\usepackage[mathscr]{eucal}
\usepackage{physics}

\begin{document}
\preprint{APS/123-QED}

\title{ Quantum signatures of bistability and limit cycle in Kerr-modified cavity magnomechanics}
\author{Pooja Kumari Gupta}
 \affiliation{Department of Physics, Indian Institute of Technology Guwahati, Guwahati-781039, Assam, India}
 
\author{Subhadeep Chakraborty}%
\affiliation{Indian Institute of Science Education and Research-Kolkata, Mohanpur, Nadia-741246, India}

\author{Sampreet Kalita}
 \affiliation{Department of Physics, Indian Institute of Technology Guwahati, Guwahati-781039, Assam, India}
 
\author{Amarendra K. Sarma}
\affiliation{Department of Physics, Indian Institute of Technology Guwahati, Guwahati-781039, Assam, India}


\begin{abstract}
We study a Kerr-modified cavity magnomechanical system with a focus on its bistable regime. We identify a distinct parametric condition under which bistability appears, featuring two stable branches and one unstable branch in the middle. Interestingly, our study reveals a unique transition where the upper branch loses its stability under a sufficiently strong drive, giving rise to limit cycle oscillation. Consequently, we report a rich phase diagram consisting of both bistable and periodic solutions and study quantum correlations around them. While in the bistable regime, we find the entanglement reaching different steady state value, in the unstable regime, entanglement oscillates in time.  This study is especially important in understanding quantum entanglement at different stable and unstable points arising in a Kerr-modified cavity magnomechanical system.

\end{abstract}

\maketitle
\section{Introduction}
\label{sec:intro}

Cavity optomechanics has undergone rapid exploration
over the passed few decades, offering a unique platform
to couple light with motion ~\cite{aspelmeyer2014cavity}. Although it started with an aim to reach the motional ground state of massive macroscopic mechanical oscillators~\cite{schliesser2006radiation, schliesser2008resolved}, research now a days has shifted its focus towards the emergence of nonlinear dynamical phenomena. Significant theoretical work has been devoted to understanding the phase space structure of classical nonlinear optomechanical systems, revealing complex behaviors including multistability~\cite{aldana2013equivalence}, limit cycles~\cite{marquardt2006dynamical, ludwig2008optomechanical, lorch2014laser, schulz2016optomechanical, roque2020nonlinear}, and chaotic dynamics~\cite{bakemeier2015route, lu2015pt}. Though experimental studies are relatively rare, they have observed crucial phenomena such as limit cycles~\cite{kippenberg2005analysis, metzger2008self}, period doubling, chaos~\cite{carmon2007chaotic, navarro2017nonlinear, das2023instabilities}, and multistable attractor diagrams~\cite{krause2015nonlinear, buters2015experimental}. Moreover, there have been attempts to explore coupled
multiple optomechanical limit cycle oscillators to study synchronization dynamics~\cite{heinrich2011collective, zhang2012synchronization, weiss2016noise}. On the quantum side, optomechanical dynamics has lead to the observation of dissipative phase transitions and bifurcation like behaviour, even with small nonlinear interactions and strong external drives \cite{ghobadi2011quantum, bibak2023dissipative, wang2024quantum}. This rich array of nonlinear behaviours and phase transitions in optomechanical systems continues to drive theoretical and practical advances in the field.

Cavity magnonics~\cite{rameshti2022cavity, yuan2022quantum}, on the other hand, has become an emerging field exhibiting strong-coupling cavity QED effects with cavity photons and magnons~\cite{huebl2013high, tabuchi2014hybridizing, zhang2014strongly}. In its simplest form, such a system comprises of an optical or microwave cavity coupled to a ferromagnetic insulator (preferably YIG), featuring high spin density and low damping rate. As the magnetic dipole coupling between the cavity photon and magnon can reach the strong coupling regime~\cite{huebl2013high, tabuchi2014hybridizing, zhang2014strongly, bai2015spin}, cavity magnonics finds potential applications in quantum information processing~\cite{osada2016cavity, hisatomi2016bidirectional, zhang2015magnon}, acting as either a transducer or memory. Besides, akin to cavity optomechanical systems, magnetostriction provides an alternate root to couple magnon with phonon~\cite{zhang2016cavity}. This has sparked a significant interest on such cavity magnomechanical system which includes phenomena like magnomechanically induced transparency~\cite{zhang2016cavity}, magnon induced dynamical backaction~\cite{potts2021dynamical}, magnomechanical squeezing~\cite{li2019squeezed} and entanglement~\cite{li2018magnon}, phonon laser~\cite{ding2019phonon}, magnon chaos~\cite{peng2024ultra}, and so on.  

Only recently has bistability in mechanical vibrations been experimentally demonstrated in a Kerr-modified cavity magnomechanical system~\cite{shen2022mechanical}. While quantum correlations in the bistable regime of optomechanical systems have been well studied~\cite{ghobadi2011quantum, bibak2023dissipative}, their counterparts in cavity magnomechanical systems still remain unexplored. Additionally, it has been established that optical bistability in optomechanical systems could exhibit behaviors similar to those observed in Kerr media~\cite{aldana2013equivalence}. However, the fluctuations in the position of the mechanical mode can destabilize the upper branch in optomechanical systems. This motivates us to study a Kerr-modified cavity magnomechanical system, with a focus on its bistable regime. Notably, bistability and squeezing have been studied recently in a nonlinear cavity magnonic system~\cite{yang2021bistability}. Our work not only recovers bistability but also uncovers a novel oscillatory macroscopic phase of magnon. In the following, we study the behaviour of the quantum fluctuations and correlations around such bistable and oscillatory phases.

The paper is organized as follows. In Sec.~\ref{sec:Model} we introduce a Kerr modified cavity magnonmechanical system that comprises a cavity mode, a magnon mode, and a phonon mode and write the quantum Langevin equations. Sec.~\ref{sec:bistability} analyses the steady states and predicts bistability in the magnon mode. In Sec.~\ref{sec: Quantum_fluctuations}, we study the dynamics of quantum fluctuations and analyse the phase space distribution of quantum correlations. Conclusion and remarks are given in Sec.~\ref{sec: conclusion}.
\newpage
\section{Model and \\ the Equations of Motion}
\label{sec:Model}

    We start with a typical cavity magnomechanical system whose Hamiltonian reads as
    \begin{align}\label{H_CMM}
        H_{CMM} &= \omega_a a^\dagger a + \omega_m m^\dagger m + \omega_b b^\dagger b + g_{ma} \left(a^\dagger m + m^\dagger a \right)\nonumber \\ 
                & + g_{mb} m^\dagger m \left(b + b^\dagger\right) + i\Omega\left(m^\dagger e^{-i\omega_d t} - m e^{i\omega_d t}\right).
    \end{align}
    Here, the first three terms of the Hamiltonian~\eqref{H_CMM}  correspond to the free energies of the cavity mode, the magnon mode and the phonon mode, respectively. The cavity mode is characterized by an annihilation (creation) operator $a$ ($a^\dagger$) and a resonance frequency (decay rate) $\omega_a$ ($\kappa_a$). The annihilation (creation) operator and the resonance frequency (decay rate) of the magnon mode are respectively given by $m$ ($m^\dagger$) and $\omega_m$ ($\kappa_m$), while the phonon mode is described by the annihilation (creation) operator $b$ ($b^\dagger$) and a resonance frequency (decay rate) $\omega_b$ ($\kappa_b$). The fourth term here corresponds to the magnetic-dipole interaction between the cavity mode and the magnon mode, with the coupling strength defined as $g_{ma}$. The coupling between the magnon and the phonon mode is given in the fifth term, described by a radiation-pressure like interaction of strength $g_{mb}$. Finally, the last term accounts for the external drive applied to the magnon mode where the drive amplitude $\Omega$ relates to the driving power $P_d$ and frequency $\omega_d$ as $\Omega = \sqrt{\frac{2\kappa_m P_d}{\hbar \omega_d}}$. 
    
    When the drive is sufficiently strong, the magnon mode further includes a Kerr-type nonlinear interaction, arising due to the magnetocrystalline anisotropy~\cite{shen2022mechanical}. The total Hamiltonian describing such a Kerr-modified cavity magnomechanical system is then given by,
    \begin{equation}
        H = H_{CMM} + K m^\dagger m m^\dagger m,
    \end{equation}
    with $K$ being the Kerr coefficient. In a rotating frame of the external drive, the dynamics of the system reads as
    \begin{subequations}
    \label{eqn:QLE}
    \begin{eqnarray}
            \dot{a}& = & -(i \Delta_{a} + \kappa_{a})a - i g_{ma} m + \sqrt{2\kappa_{a}} a^{in},
            \\
             \dot{m} & = & -(i \Delta_{m} + \kappa_{m})m - i g_{ma} a -2 i K m^{\dagger} m m
             \\
             && - i g_{mb} m (b + b^{\dagger}) +\Omega + \sqrt{2\kappa_{m} }m^{in}, \nonumber
             \\
            \dot{b} & = & -(i \omega_{b} + \kappa_{b})b - i g_{mb}m^{\dagger} m +  \sqrt{2\kappa_{b} }b^{in}, 
    \end{eqnarray}
    \end{subequations}
    where $\Delta_a = \omega_a - \omega_d$ and $\Delta_m = \omega_m - \omega_d$ respectively define the cavity and the magnon detuning, and $a_{in}$, $m_{in}$, and $b_{in}$ are the input quantum noise operators, respectively associated with the cavity, magnon, and the phonon modes. With zero-mean values, the noise operators are characterized by the following correlation functions: $\langle a^{in}(t) a^{in,\dagger} (t^\prime)\rangle = \delta(t-t^\prime)$, $\langle m^{in}(t) m^{in,\dagger} (t^\prime)\rangle = \delta(t-t^\prime)$, $\langle b^{in}(t) b^{in,\dagger} (t^\prime)\rangle = (n_{th}+1)\delta(t-t^\prime)$ and 
    $\langle b^{in,\dagger}(t) b^{in} (t^\prime)\rangle = n_{th}\delta(t-t^\prime)$, where $n_{th}=\left[e^{\left(\frac{\hbar \omega_b}{K_B T}\right)}-1\right]^{-1}$ denotes the number of thermal phonon at temperature $T$ and $K_B$ is the Boltzmann constant.

\section{Steady States and \\Magnon Bistability}
\label{sec:bistability}
    For a strongly driven magnon mode, each operator $\mathcal{O}$ ($\mathcal{O}=a,m,b$) is assumed to reach a steady state ($\mathcal{\dot{O}}=0$), characterized by a large mean value $\langle \mathcal{O}\rangle\gg 1$. Following the substitution $\mathcal{O}$=$\langle \mathcal{O} \rangle$, with the left-hand side of the Eq.~\eqref{eqn:QLE} set to zero, the steady state (fixed point) values are given by 
    \begin{eqnarray}
        \label{eqn:mean_val}
             \langle a\rangle & = & - \frac{g_{ma} \langle m\rangle }{ (\Delta_{a} - i \kappa_{a})}  \nonumber
            \\
             \langle m \rangle & = - & \frac{g_{ma} \langle a \rangle}{\Delta_{m}^\prime - i \kappa_{m}} - \frac{i \Omega}{\Delta_m^\prime - i\kappa_m}\nonumber
            \\
            \langle b \rangle & = & - \frac{g_{mb} \lvert\langle  m \rangle\rvert^2}{(\omega_{b} - i \kappa_{b})},
    \end{eqnarray}
    where $\Delta_m^\prime = \Delta_{m} + 2K {\lvert\langle m \rangle\rvert}^2 + 2 g_{mb}\mathrm{Re}[\langle b \rangle]$ represents the effective magnon detuning which includes the frequency shift, arising due to the combined effect of self-Kerr nonlinearity and the magnomechanical coupling. Note that in Eq.~\eqref{eqn:mean_val}, we have decomposed the product operators as $\langle A B \rangle = \langle A \rangle \langle B \rangle$ which is valid in the semi-classical limit $\langle\mathcal{O}\rangle\gg 1$. In what follows, we solve Eq.~\eqref{eqn:mean_val} and obtain a third order nonlinear equation 
    \begin{equation}
        \label{eqn:bistability}
        K^{\prime 2} I^{3} + 2 \Delta_{0}  K^{\prime} I^{2} + (\Delta_{0}^{2} + \kappa_{0}^{2}) I - \Omega^{2}  =  0,
    \end{equation}
    where $I ={\lvert\langle m \rangle\rvert}^2$ defines the mean magnon number and $K^{\prime} = 2(K - \zeta \omega_{b} )$ is the effective Kerr-nonlinearity. The other parameters are defined as follows: $\Delta_{o}= \Delta_{m}- \eta \Delta_{a}$, $\kappa_{o}= \kappa_{m} + \eta \kappa_{a}$, $\eta= g_{ma}^{2}/\bigl( \Delta_{a}^{2} + \kappa_{a}^{2} \bigr)$ and $\zeta= g_{mb}^{2}/\bigl( \omega_{b}^{2} + \kappa_{b}^{2} \bigr)$. The cubic Eq.~\eqref{eqn:bistability} can have one or three real roots depending on the parameters $K^\prime$, $\Omega$, $\Delta_0$ and $\kappa_0$. We find that Eq.~\eqref{eqn:bistability} possess three distinct roots only if the discriminant of the equation satisfies
    \begin{multline}
    \label{eqn: three_roots}
        27 K^{\prime 2} \Omega^{4} + 4 \Delta_{0} K^{\prime} \Omega^{2}(\Delta_{o}^{2} + 9\kappa_{o}^{2}) \\
        +4 \kappa_{0}^{2}(\Delta_{o}^{2} + \kappa_{o}^{2})^{2} < 0.
    \end{multline}
    Upon fulfilment of Eq.~\eqref{eqn: three_roots}, the solutions of Eq.~\eqref{eqn:bistability} form the characteristic \textit{S}-shaped curve with two switching points at which the derivative $\frac{d\Omega}{dI}=0$, i.e.,
    \begin{equation}
    \label{eqn:quadratic}
        3 K^{\prime 2 } I^{2} + 4 \Delta_{0}  K^{\prime} I + (\Delta_{0}^{2} + \kappa_{0}^{2}) =  0.
    \end{equation}
    The quadratic equation above has two real roots, representing the two switching points, if and only if the discriminant of Eq.~\eqref{eqn:quadratic} satisfies
    \begin{equation}
    \label{eqn:delta_0_K_O}
    \Delta_{0}^{2} - 3 \kappa_{0}^{2} > 0.
    \end{equation}
    While at $\Delta_{0}^{2}- 3 \kappa_{0}^{2}=0$, the equation yields two real and identical roots. This is to say that at $\Delta_0=\pm\sqrt{3}\kappa_0$ the two switching points merge into one, implying an absence of bistability. Consequently, we derive a critical driving strength $\Omega_{c} = \sqrt{\frac{-8 \Delta_{0}^{3}}{27 K^{\prime}}}$ above which bistability appears. Since the critical drive amplitude $\Omega_{c}$ cannot be negative, one has to have a positive $K^{\prime}$ with negative $\Delta_{0}$ or vice versa. Upon further simplification, the bistability condition gets reduced to $\Delta_{m} < \eta \Delta_{a}$(for $K^{\prime}>0$) or $\Delta_{m} > \eta \Delta_{a}$ (for $K^{\prime}<0$).

    \begin{figure}[t]
        \centering
        \includegraphics[width=0.84\textwidth]{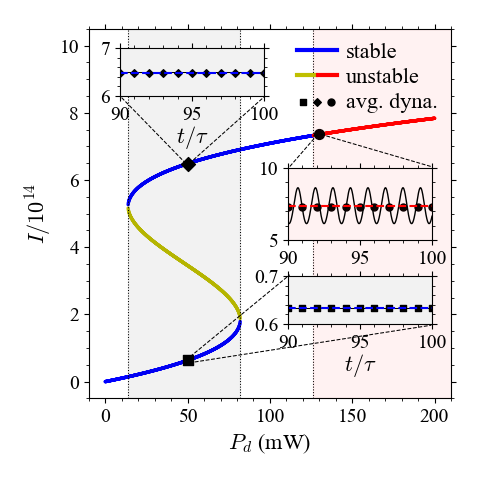}
        \caption{Magnon number $\textit{I} = |\langle m \rangle|^{2}$ versus the driving power $P_d$. The insets at $\blacksquare$, $\blacklozenge$ and \begin{tikzpicture}[scale=0.85]\fill (0,0) circle (3pt);\end{tikzpicture} respectively correspond to the long-time dynamics at the lower-stable, upper-stable and upper-unstable fixed points.
        For simulation, we use $\omega_{a}/2\pi = 10$GHz, $\omega_{b}/2\pi = 10$ MHz, $\kappa_{a}/2\pi = \kappa_{m}/2\pi = 1$ MHz, $\kappa_{b}/2\pi=100$ Hz, $\Delta_{a} = -0.9\omega_{b}$, $\Delta_{m} = -0.8\omega_{b}$, $K/2\pi = 6.5$ nHz, $g_{ma}/2\pi = 3.2$ MHz, $g_{mb}/2\pi = 1$ mHz, $\tau = 2\pi/\omega_b$, $T = 10$mK.}
        \label{fig:bistability}
    \end{figure}
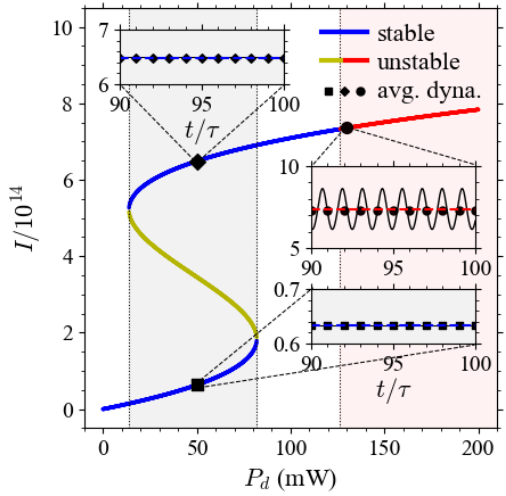

   Next, to assess the stability of these steady states, we consider small fluctuations $\delta \mathcal{O}$ around the fixed points and rewrite each operator as: $\mathcal{O} = \langle \mathcal{O} \rangle + \delta \mathcal{O}$. The linearized dynamics of the fluctuations is then given by
    \begin{equation}
        \label{eqn: fluctuation_dynamics}
            \dot{u}(t) = A(t) u(t) + \xi(t),
    \end{equation}
    where $u(t)= \left[ \delta X_{a}, \delta Y_{a}, \delta X_{m}, \delta Y_{m},\delta X_{b}, \delta Y_{b}\right]^{T}$ and $\xi(t) =\left[\sqrt{2 \kappa_{a}} X^{in}_{a},\sqrt{2 \kappa_{a}} Y^{in}_{a}, \sqrt{2 \kappa_{m}} X^{in}_{m}, \sqrt{2 \kappa_{m}} Y^{in}_{m}, \sqrt{2 \kappa_{b}} X^{in}_{b},\right. \\ \left.
    \sqrt{2 \kappa_{b}} Y^{in}_{b}\right]^{T}$  are respectively the vector of fluctuations and input noises, with their quadratures defined as $\delta X_{\mathcal{O}}= (\delta \mathcal{O}  + \delta \mathcal{O}^{\dagger})/\sqrt{2}$, $\delta Y_{\mathcal{O}}= (\delta \mathcal{O}  - \delta \mathcal{O}^{\dagger})/\sqrt{2}i$ and $\delta X^{in}_{\mathcal{O}}= (\delta \mathcal{O}^{in}  + \delta \mathcal{O}^{in,\dagger})/\sqrt{2}$, $\delta Y_{\mathcal{O}}^{in}= (\delta \mathcal{O}^{in}  - \delta \mathcal{O}^{in,\dagger})/\sqrt{2}i$ ($\mathcal{O}=a,m,b$). The matrix $A$ reads as $A(t)=$
    \begin{equation}
        \label{eqn:drift_matrix}
            \begin{pmatrix}
            -\kappa_{a} & \Delta_{a} & 0 & g_{ma} & 0 & 0 &\\[6pt]
            
            -\Delta_{a} & -\kappa_{a }& -g_{ma} & 0 & 0 & 0 \\[6pt]
            
            0 & g_{ma} & -\kappa_{m} + \Delta^y_K & \Delta^{\prime\prime}_{m}-\Delta^x_K & G^y_{mb} & 0\\[6pt]
            
            -g_{ma} & 0 & -\Delta^{\prime\prime}_{m}-\Delta^x_K & -\kappa_{m}- \Delta^y_K & -G^x_{mb} & 0\\[6pt]
            
            0 & 0 & 0 & 0 & -\kappa_{b} & \omega_{b}\\[6pt]
            
            0 & 0 & -G^x_{mb} & -G^y_{mb} & -\omega_{b} & -\kappa_{b}
        \end{pmatrix},
    \end{equation}
    with the coefficients being given by the following expressions: $\Delta^{\prime \prime}_m = \Delta_m^\prime + 2K\lvert\langle m \rangle \rvert^2$, $\Delta_K = 2K\langle m \rangle^2 = \Delta^x_K + i \Delta^y_K$, and $G_{mb} = 2 g_{mb} \langle m \rangle = G^x_{mb} + i G^y_{mb}$. The system is said to be stable when all the eigenvalues of $A$ evaluated at the steady states have negative real parts.

   \begin{figure}[t]
        \centering
        \includegraphics[width=0.84\textwidth]{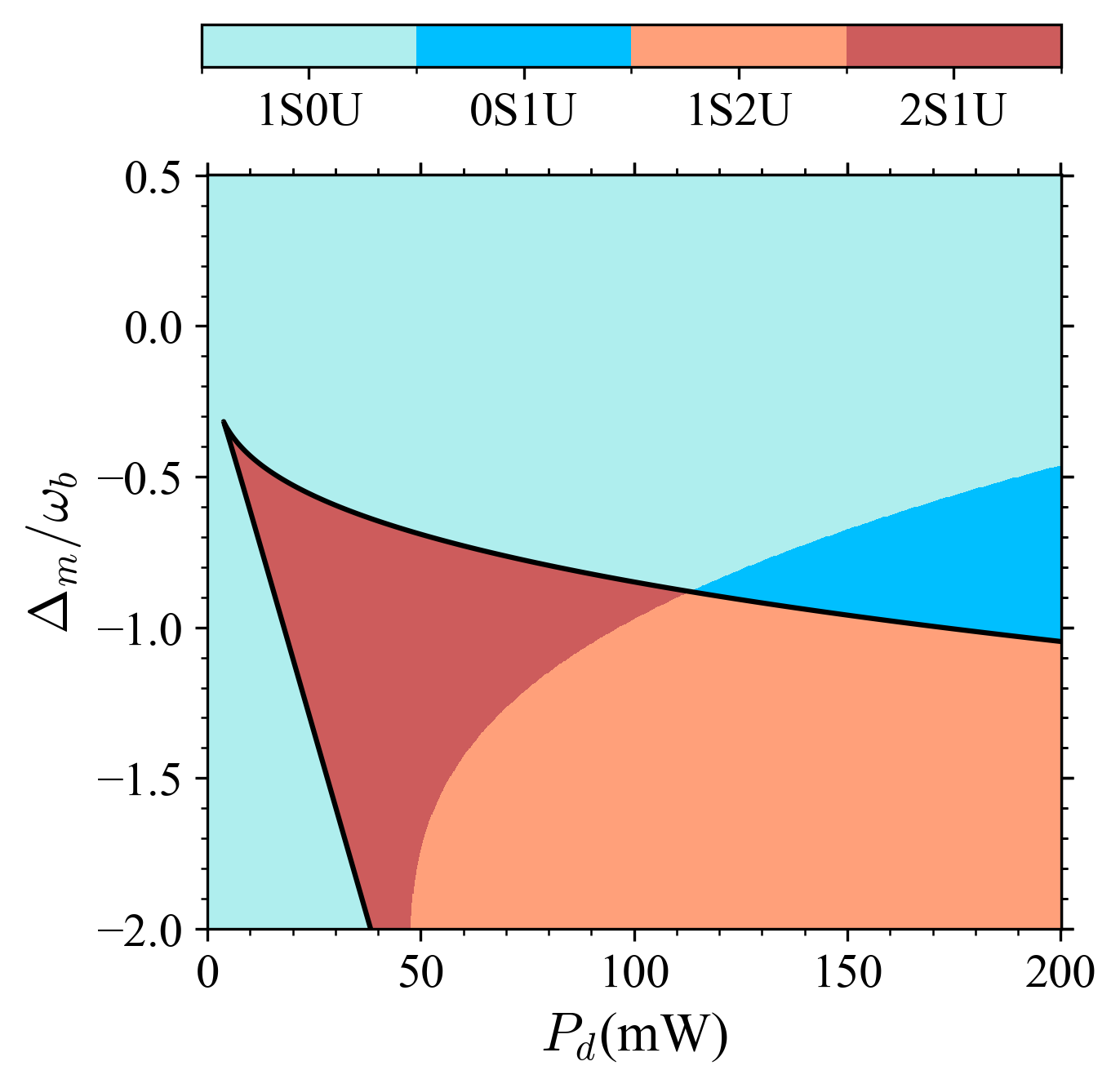}
        \caption{Behaviour of the steady state solutions as a function of the driving power $P_d$ and the magnon detuning $\Delta_m$. The region bounded by the black line contains three different solutions. In region $\mathrm{2S1U}$, two of them are stable, while in $\mathrm{1S2U}$, only one of them is stable. The rest of the regions consist a single solution that is either stable $\mathrm{1S0U}$ or unstable $\mathrm{0S1U}$. All the parameters remain same as Fig.~\ref{fig:bistability}.}
        \label{fig:phase_diagram}
    \end{figure}

    In Fig.\ref{fig:bistability} we plot the mean magnon number with respect to the driving power $P_d$. Notice that in the range $13.84\leq P_d\leq 81.94$ the solutions of Eq.~\eqref{eqn:bistability} form the characteristic \textit{S}-shaped curve, with two stable (the lower and the upper) and one unstable (middle) branch. This is followed by a single steady state solution which changes its stability at a critical power $P^c_d=126.34$~mW. Interestingly, the type of instabilities found in the upper and the middle branches are quite different. To demonstrate this difference, we plot the dynamics of the mean magnon number at different powers. We first consider $P_d = 50$~mW where the system exhibits bistability. The insets at $\blacksquare$ and $\blacklozenge$ show that the magnon mode can settle in either a low or a high value depending on its initial condition. Notably, we find the basin of attractions for these two steady states are well separated. Fixing $P_d$ at 130~mW (\begin{tikzpicture}[scale=0.85]\fill (0,0) circle (3pt);\end{tikzpicture}), we observe an onset of oscillatory solution. Such oscillations essentially yields limit-cycle trajectory when looked into its phase-space dynamics. Notably, here the eigenvalues of $A(t)$ appear as complex conjugate pairs, a characteristic feature of Hopf bifurcation.

    Fig.\ref{fig:phase_diagram} depicts the solution of the cubic Eq.~\eqref{eqn:bistability} in $(P,\Delta_m$) plane. We identify four different categories of solutions based on their number and stability. These are as follows: one stable fixed point ($\mathrm{1S0U}$), two stable and one unstable fixed point ($\mathrm{2S1U}$), one unstable fixed point ($\mathrm{0S1U}$), and one stable and two unstable fixed points ($\mathrm{1S2U}$).
    One distinctive feature of our model is that bistability appears in the blue sideband regime. This is further confirmed from the analysis followed by Eq.~\eqref{eqn:delta_0_K_O}. We next focus on the behaviour of quantum entanglement at these distinctive phases of stable and unstable fixed points.

    \begin{figure}[t]
        \centering
        \includegraphics[width=0.84\textwidth]{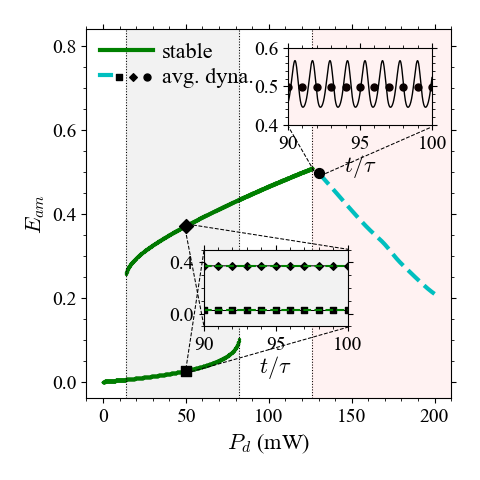}
        \caption{Magnon-photon entanglement $E_{am}$ against the driving power $P_{d}$. The steady state and the time-averaged entanglement are respectively delineated by the (green) solid and (cyan) dashed line. The long time dynamics $E_{am}$ for the bistable ($\blacksquare$ and $\blacklozenge$) and the unstable (\begin{tikzpicture}[scale=0.85]\fill (0,0) circle (3pt);\end{tikzpicture}) fixed points are respectively shown in the lower-middle and top-right insets. We keep all the parameters same as Fig.~\ref{fig:bistability}}     
        \label{fig:E_am_bistability}
    \end{figure}
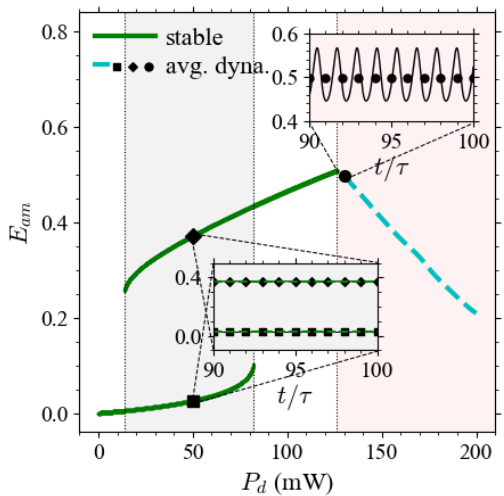

\section{Quantum Correlation \\ and fluctuations dynamics}
\label{sec: Quantum_fluctuations}
    Due to the linearized dynamics and zero-mean Gaussian nature of the quantum noises, the tripartite quantum system can be fully characterized by a $6\times6$ correlation matrix (CM), with its elements defined as $ V_{ij} = \langle u_{i}(t)u_{j}(t) + u_{j}(t)u_{i}(t) \rangle /2$. The dynamics of the CM~\cite{sarma2021continuous} can then be derived as
    \begin{equation}
        \label{eqn:cov_dynamics}
        \dot{V}(t) = A(t)V(t) + V(t) A^{T}(t) + D,
    \end{equation}
    where $D = \mathrm{diag}( \kappa_{a}, \kappa_{a}, \kappa_{m}, \kappa_{m}, \kappa_{b}(2 n_{th} + 1), \kappa_{b}(2 n_{th} + 1) )$ is the diffusion matrix. Note that Eq.~\eqref{eqn:cov_dynamics} is an inhomogeneous first-order differential equation that can be numerically solved starting from an initial condition. In what follows, we assume that the cavity and the magnon modes are prepared in a coherent state while the phonon mode is in a thermal state at a temperature $T$.

    To estimate entanglement and other related quantities, we extract the CM comprising the cavity and the magnon mode. Such a bipartite system~\cite{adesso2014continuous, sarma2021continuous} is commonly expressed as
    \begin{equation}
     \label{eqn: reduced_Vam}
     V^{(2)}\equiv
     \begin{pmatrix}
      \alpha & \beta\\
      \beta^{T} & \gamma
     \end{pmatrix},
    \end{equation}
    where $\alpha$, $\gamma$ and $\beta$ are $2\times2$ matrices, respectively characterizing the cavity mode, the magnon mode, and the correlation between them. The degree of bipartite entanglement can then be quantified by the so-called logarithmic negativity~\cite{vidal2002computable}, defined as 
     $E_{N} \equiv \mathrm{max}[0, -\ln(2 \nu^{-})]$, where $\nu^{-} \equiv 2^{-1/2}[ \Sigma(V^{(2)}) - \sqrt{ \Sigma(V^{(2)})^{2} - 4 \det V^{(2)}]}^{1/2}$ and $\Sigma(V^{(2)}) = \det(\alpha) + \det(\gamma)-2\det(\beta)$.

    Fig.\ref{fig:E_am_bistability} depicts the entanglement $E_{am}$ between the cavity and the magnon mode against the driving power $P_d$. Notably the signature of bistability is also evident in the entanglement behaviour. It is shown that when following the lower branch of the bistability curve, the cavity and the magnon modes get entangled which becomes maximum at the switching point where the lower branch becomes unstable. Along the upper branch, the entanglement remains noticeably high, reaching its maximum at the critical power $P^c_d$ (where the upper branch loses its stability).
    The inset combining $\blacksquare$ and $\blacklozenge$ depicts the corresponding entanglement dynamics. Starting from an initially unentangled state, the dynamics quickly saturates to respective steady state values corresponding to the lower and upper (stable) branches of the bistability curve. While in the Hopf bifurcation regime where no stable fixed points exist, the entanglement oscillates in time (\begin{tikzpicture}[scale=0.85]\fill (0,0) circle (3pt);\end{tikzpicture}). We note that such a dynamical behaviour stems from the oscillatory solution of the (large) semiclassical mean values. However, going deep into the unstable regime, we observe the formation of a beat-like pattern with a rapid decrease in the mean oscillation amplitude. The time-averaged entanglement at different driving amplitude is shown by the (cyan) dashed line in Fig.~\ref{fig:E_am_bistability}.
    
\begin{figure*}[t]
        \centering
        \includegraphics[width=0.84\textwidth]{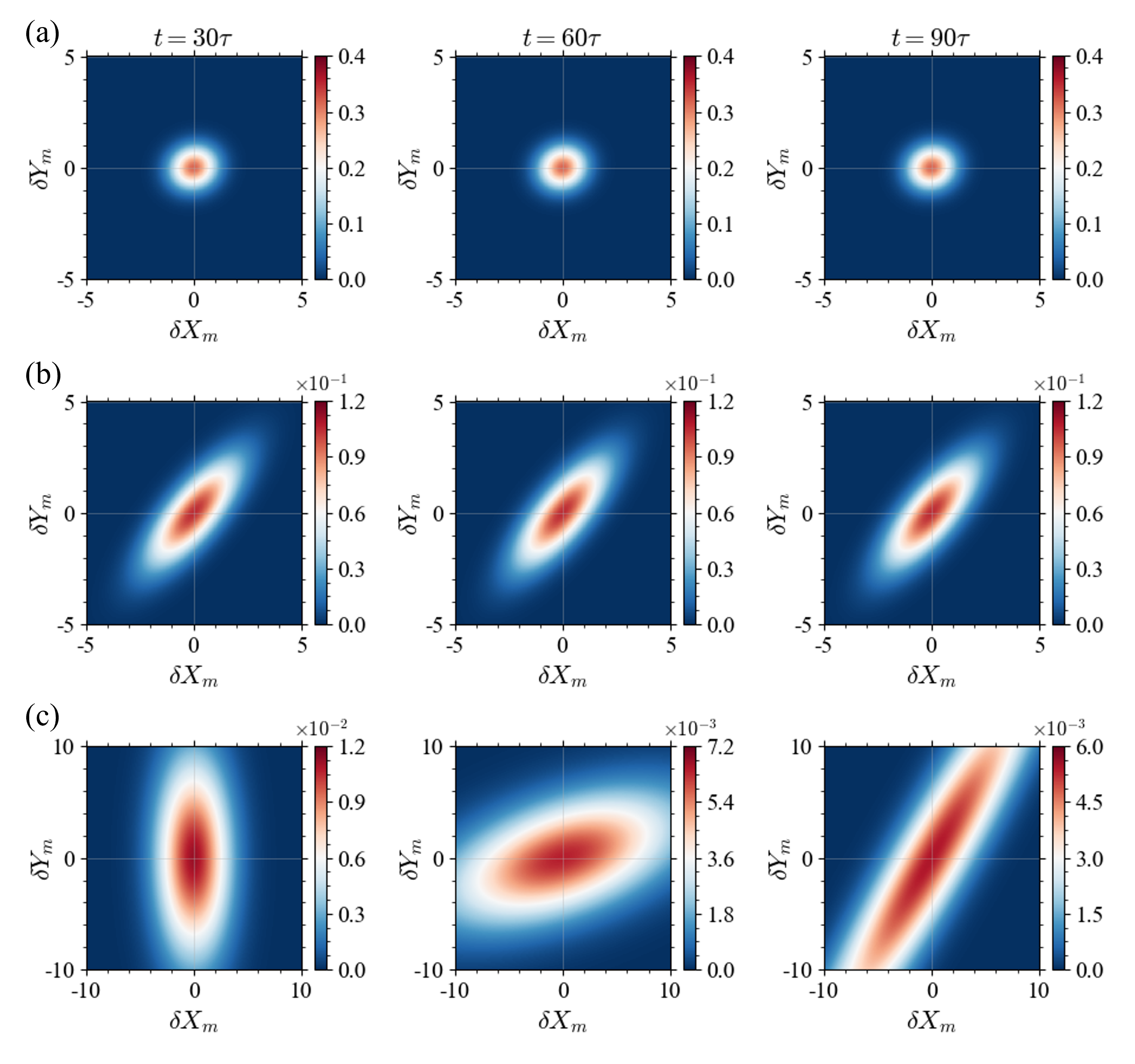}
        
        \caption{ Wigner distribution of the magnon mode at different instant of times $t=30\tau, 60\tau$, and $90\tau$, The top (a), middle (b) and the bottom (c) panel respectively correspond to the two stable $\blacksquare$ (lower), $\blacklozenge$ (upper), and one unstable \begin{tikzpicture}[scale=0.85]\fill (0,0) circle (3pt);\end{tikzpicture} points,
        as marked in the Fig.\ref{fig:bistability}. }
        \label{fig:wigner}
    \end{figure*}
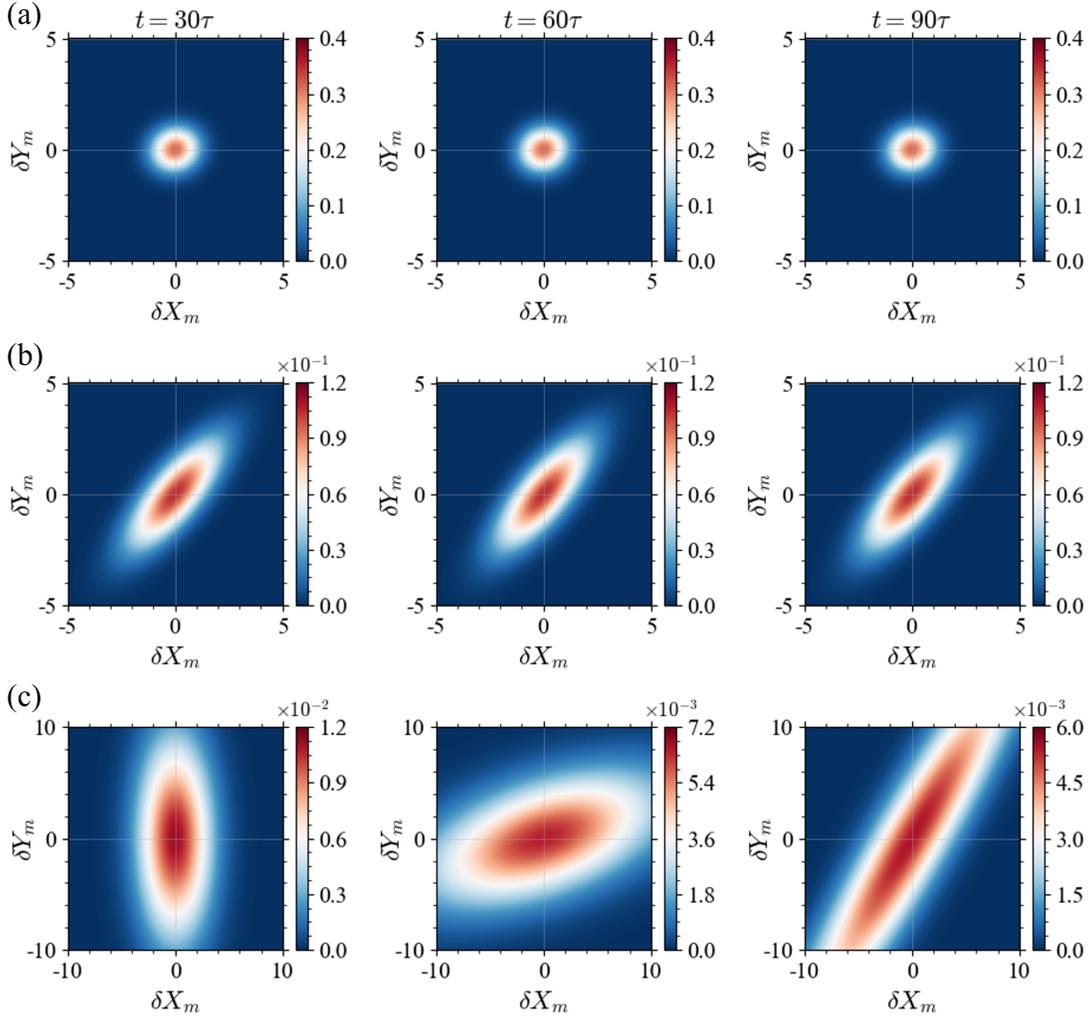

     To understand the emergent dynamical instabilities of the entanglement, we next look into the phase-space dynamics of the quantum fluctuations. We particularly consider the Wigner distribution~\cite{adesso2014continuous, sarma2021continuous} of the magnon mode 
    \begin{equation}
        \label{wigner}
        \mathcal{W}(u_m)=\frac{1}{2\pi\mathrm{det}[\gamma]}\mathrm{exp}\left[-\frac{1}{2}u^T_m\gamma^{-1}u_m\right],
    \end{equation}
    where the state vector reads as $u_m=\left[ \delta X_{m}, \delta Y_{m}\right]^{T}$ and plot it at different instant of time. It is seen that for values of the driving power at the bistable regime ($\blacksquare$, $\blacklozenge$), both the lower  (Fig. \ref{fig:wigner}(a)) and upper (Fig. \ref{fig:wigner}(b)) branches of the bistability curve remain squeezed. However, one finds a relatively higher degree of squeezing in the upper branch. Of more importance though, we find the Wigner functions being localized in the phase space for both these stable branches. While focusing at the unstable regime (\begin{tikzpicture}[scale=0.85]\fill (0,0) circle (3pt);\end{tikzpicture}), the Wigner function reveals an interesting behaviour (Fig. \ref{fig:wigner}(c)). We find that during its evolution the Wigner function rotates and gets stretched along its anti-squeezing axis. This stretching could be corroborated to the arsing instabilities in the magnon mode induced by blue-sideband drive~\cite{ludwig2008optomechanical}.

    In passing, we note that, unlike existing proposals for cavity magnomechanical entanglement~\cite{li2018magnon, yu2020magnetostrictively, li2019entangling, li2021entangling}, we achieve cavity-magnon entanglement using experimentally demonstrated parameters~\cite{zhang2016cavity,shen2022mechanical}. However, no entanglement is observed between the cavity-phonon and magnon-phonon modes due to the low magnomechanical coupling strength. In our system, the entanglement $E_{am}$ primarily originates from the Kerr nonlinear term, as the low $g_{mb}$ does not contribute to entanglement formation. A key distinction emerges when comparing entanglement behavior along the upper branch of optomechanical and magnomechanical systems. In optomechanical systems, entanglement decreases monotonically along the upper branch of the bistability curve~\cite{ghobadi2011quantum}. In contrast, in magnomechanical systems, entanglement peaks until the upper branch reaches instability (Fig.~\ref{fig:E_am_bistability}). This difference could be attributed to the distinct stability landscapes that arise in each system under the chosen parameter regime.

\begin{figure}[t]
    \centering
    \includegraphics[width=0.84\linewidth]{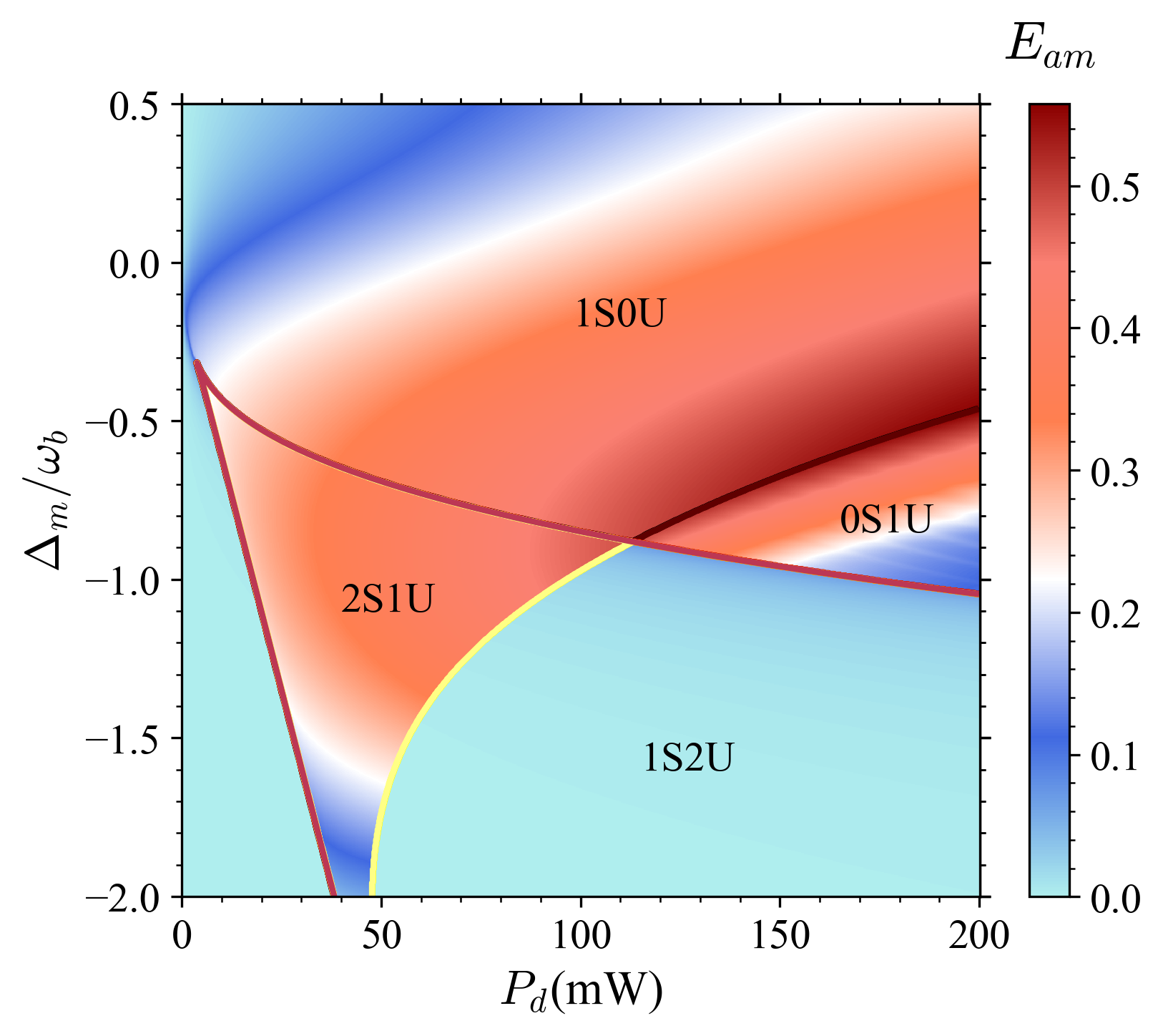}
    \caption{Phase diagram of magnon-photon entanglement $E_{am}$ versus $P_{d}$ and  detuning $\Delta_{m}$. The parameters are kept same as those in Fig.~\ref{fig:bistability}.}
    \label{fig:Eam_phase}
    \end{figure}

     Next, to illustrate the variation of magnon-photon entanglement across different stability regions, in Fig. \ref{fig:Eam_phase} we plot the phase diagram of $E_{am}$ in $(P_d,\Delta_m)$ plane. It is observed that the global maxima of entanglement occurs at the boundary between $\mathrm{1S0U}$ and $\mathrm{0S1U}$, i.e. where a single stable solution becomes unstable. Notably, this can be considered a strong quantum fingerprint of the transition from stable fixed points to limit cycle~\cite{meng2020quantum}. While in the bistable region $\mathrm{2S1U}$, where two stable branches exist, we focus only on the upper branch, highlighting the maximum degree of quantum entanglement. Conversely, in the 1S2U region, hosting a single stable solution, the entanglement is found to be quite low. This is due to the weak effective Kerr-nonlinear term contributing to the entanglement formation. At $\mathrm{0S1U}$ where the system losses stability, we show the time-averaged entanglement value. In agreement to Fig.~\ref{fig:E_am_bistability}, we see rapid decrease of mean entanglement as we move away from the boundary.

    Finally, we comment on the considered strong driving limit and the linearized description. In principle, when the parameters are set to the bistable regime, one naively expect the Wigner distribution to exhibit a bimodal structure~\cite{aldana2013equivalence}. Likewise, the existence of limit cycle should correspond to a ring-shaped Wigner distribution~\cite{lorch2014laser, amitai2017synchronization}. However, this requires a full simulation of the quantum master equation, demanding strong nonlinearity and weak driving. In particular, to our system this becomes practically intractable due to large size of the Hilbert space, associated with the combined cavity-magnon-phonon modes. Moreover, the presence of external noise makes the problem even more intriguing. It is expected that due to this noises, the phase-space distribution of fixed points gets smeared out. However, in the low temperature limit, the noises has minimal influence, leading to longer transient times before the quantum correlations vanishes abruptly. Nevertheless, these still remain open questions and deserve further study.

\section{Conclusion}
\label{sec: conclusion}
    We studied a Kerr-modified cavity magnomechanical system. We recovered bistability in mean magnon number. Apart from bistability, our study reveals that the upper stable branch undergoes a Hopf bifucation and becomes unstable at a relatively high driving power. We subsequently studied the behaviour of quantum entanglement around these stable and unstable fixed points. While we observed the signatures of bistability in the quantum entanglement, a novel oscillatory behaviour was found around the unstable fixed point. Interestingly, the entanglement
    reaches its maximum at the boundary between the stable and unstable fixed points. We further note that throughout the entire parameter space, our study does not lead to chaos as confirmed through a negative Lyapunov exponent. As chaos has been recently found in a Kerr-modified cavity magnomechanical system~\cite{peng2024ultra}, we believe that the exploration of nonlinear dynamics alongside quantum entanglement would be a worthwhile study~\cite{wang2014nonlinear}.

\section*{Acknowledgement}
P.K.G. gratefully acknowledges a research fellowship from MoE, Government of India. A.K.S.  acknowledges the grant from MoE, Government of India (Grant No. MoE-STARS/STARS- 2/2023-0161).

\bibliography{bistability_CMM_ref}

\providecommand{\noopsort}[1]{}\providecommand{\singleletter}[1]{#1}%
\begin{thebibliography}{50}%
\makeatletter
\providecommand \@ifxundefined [1]{%
 \@ifx{#1\undefined}
}%
\providecommand \@ifnum [1]{%
 \ifnum #1\expandafter \@firstoftwo
 \else \expandafter \@secondoftwo
 \fi
}%
\providecommand \@ifx [1]{%
 \ifx #1\expandafter \@firstoftwo
 \else \expandafter \@secondoftwo
 \fi
}%
\providecommand \natexlab [1]{#1}%
\providecommand \enquote  [1]{``#1''}%
\providecommand \bibnamefont  [1]{#1}%
\providecommand \bibfnamefont [1]{#1}%
\providecommand \citenamefont [1]{#1}%
\providecommand \href@noop [0]{\@secondoftwo}%
\providecommand \href [0]{\begingroup \@sanitize@url \@href}%
\providecommand \@href[1]{\@@startlink{#1}\@@href}%
\providecommand \@@href[1]{\endgroup#1\@@endlink}%
\providecommand \@sanitize@url [0]{\catcode `\\12\catcode `\$12\catcode `\&12\catcode `\#12\catcode `\^12\catcode `\_12\catcode `\%12\relax}%
\providecommand \@@startlink[1]{}%
\providecommand \@@endlink[0]{}%
\providecommand \url  [0]{\begingroup\@sanitize@url \@url }%
\providecommand \@url [1]{\endgroup\@href {#1}{\urlprefix }}%
\providecommand \urlprefix  [0]{URL }%
\providecommand \Eprint [0]{\href }%
\providecommand \doibase [0]{https://doi.org/}%
\providecommand \selectlanguage [0]{\@gobble}%
\providecommand \bibinfo  [0]{\@secondoftwo}%
\providecommand \bibfield  [0]{\@secondoftwo}%
\providecommand \translation [1]{[#1]}%
\providecommand \BibitemOpen [0]{}%
\providecommand \bibitemStop [0]{}%
\providecommand \bibitemNoStop [0]{.\EOS\space}%
\providecommand \EOS [0]{\spacefactor3000\relax}%
\providecommand \BibitemShut  [1]{\csname bibitem#1\endcsname}%
\let\auto@bib@innerbib\@empty
\bibitem [{\citenamefont {Aspelmeyer}\ \emph {et~al.}(2014)\citenamefont {Aspelmeyer}, \citenamefont {Kippenberg},\ and\ \citenamefont {Marquardt}}]{aspelmeyer2014cavity}%
  \BibitemOpen
  \bibfield  {author} {\bibinfo {author} {\bibfnamefont {M.}~\bibnamefont {Aspelmeyer}}, \bibinfo {author} {\bibfnamefont {T.~J.}\ \bibnamefont {Kippenberg}},\ and\ \bibinfo {author} {\bibfnamefont {F.}~\bibnamefont {Marquardt}},\ }\href@noop {} {\bibfield  {journal} {\bibinfo  {journal} {Reviews of Modern Physics}\ }\textbf {\bibinfo {volume} {86}},\ \bibinfo {pages} {1391} (\bibinfo {year} {2014})}\BibitemShut {NoStop}%
\bibitem [{\citenamefont {Schliesser}\ \emph {et~al.}(2006)\citenamefont {Schliesser}, \citenamefont {Del’Haye}, \citenamefont {Nooshi}, \citenamefont {Vahala},\ and\ \citenamefont {Kippenberg}}]{schliesser2006radiation}%
  \BibitemOpen
  \bibfield  {author} {\bibinfo {author} {\bibfnamefont {A.}~\bibnamefont {Schliesser}}, \bibinfo {author} {\bibfnamefont {P.}~\bibnamefont {Del’Haye}}, \bibinfo {author} {\bibfnamefont {N.}~\bibnamefont {Nooshi}}, \bibinfo {author} {\bibfnamefont {K.}~\bibnamefont {Vahala}},\ and\ \bibinfo {author} {\bibfnamefont {T.~J.}\ \bibnamefont {Kippenberg}},\ }\href@noop {} {\bibfield  {journal} {\bibinfo  {journal} {Physical Review Letters}\ }\textbf {\bibinfo {volume} {97}},\ \bibinfo {pages} {243905} (\bibinfo {year} {2006})}\BibitemShut {NoStop}%
\bibitem [{\citenamefont {Schliesser}\ \emph {et~al.}(2008)\citenamefont {Schliesser}, \citenamefont {Rivi{\`e}re}, \citenamefont {Anetsberger}, \citenamefont {Arcizet},\ and\ \citenamefont {Kippenberg}}]{schliesser2008resolved}%
  \BibitemOpen
  \bibfield  {author} {\bibinfo {author} {\bibfnamefont {A.}~\bibnamefont {Schliesser}}, \bibinfo {author} {\bibfnamefont {R.}~\bibnamefont {Rivi{\`e}re}}, \bibinfo {author} {\bibfnamefont {G.}~\bibnamefont {Anetsberger}}, \bibinfo {author} {\bibfnamefont {O.}~\bibnamefont {Arcizet}},\ and\ \bibinfo {author} {\bibfnamefont {T.~J.}\ \bibnamefont {Kippenberg}},\ }\href@noop {} {\bibfield  {journal} {\bibinfo  {journal} {Nature Physics}\ }\textbf {\bibinfo {volume} {4}},\ \bibinfo {pages} {415} (\bibinfo {year} {2008})}\BibitemShut {NoStop}%
\bibitem [{\citenamefont {Aldana}\ \emph {et~al.}(2013)\citenamefont {Aldana}, \citenamefont {Bruder},\ and\ \citenamefont {Nunnenkamp}}]{aldana2013equivalence}%
  \BibitemOpen
  \bibfield  {author} {\bibinfo {author} {\bibfnamefont {S.}~\bibnamefont {Aldana}}, \bibinfo {author} {\bibfnamefont {C.}~\bibnamefont {Bruder}},\ and\ \bibinfo {author} {\bibfnamefont {A.}~\bibnamefont {Nunnenkamp}},\ }\href@noop {} {\bibfield  {journal} {\bibinfo  {journal} {Physical Review A—Atomic, Molecular, and Optical Physics}\ }\textbf {\bibinfo {volume} {88}},\ \bibinfo {pages} {043826} (\bibinfo {year} {2013})}\BibitemShut {NoStop}%
\bibitem [{\citenamefont {Marquardt}\ \emph {et~al.}(2006)\citenamefont {Marquardt}, \citenamefont {Harris},\ and\ \citenamefont {Girvin}}]{marquardt2006dynamical}%
  \BibitemOpen
  \bibfield  {author} {\bibinfo {author} {\bibfnamefont {F.}~\bibnamefont {Marquardt}}, \bibinfo {author} {\bibfnamefont {J.}~\bibnamefont {Harris}},\ and\ \bibinfo {author} {\bibfnamefont {S.~M.}\ \bibnamefont {Girvin}},\ }\href@noop {} {\bibfield  {journal} {\bibinfo  {journal} {Physical review letters}\ }\textbf {\bibinfo {volume} {96}},\ \bibinfo {pages} {103901} (\bibinfo {year} {2006})}\BibitemShut {NoStop}%
\bibitem [{\citenamefont {Ludwig}\ \emph {et~al.}(2008)\citenamefont {Ludwig}, \citenamefont {Kubala},\ and\ \citenamefont {Marquardt}}]{ludwig2008optomechanical}%
  \BibitemOpen
  \bibfield  {author} {\bibinfo {author} {\bibfnamefont {M.}~\bibnamefont {Ludwig}}, \bibinfo {author} {\bibfnamefont {B.}~\bibnamefont {Kubala}},\ and\ \bibinfo {author} {\bibfnamefont {F.}~\bibnamefont {Marquardt}},\ }\href@noop {} {\bibfield  {journal} {\bibinfo  {journal} {New Journal of Physics}\ }\textbf {\bibinfo {volume} {10}},\ \bibinfo {pages} {095013} (\bibinfo {year} {2008})}\BibitemShut {NoStop}%
\bibitem [{\citenamefont {L{\"o}rch}\ \emph {et~al.}(2014)\citenamefont {L{\"o}rch}, \citenamefont {Qian}, \citenamefont {Clerk}, \citenamefont {Marquardt},\ and\ \citenamefont {Hammerer}}]{lorch2014laser}%
  \BibitemOpen
  \bibfield  {author} {\bibinfo {author} {\bibfnamefont {N.}~\bibnamefont {L{\"o}rch}}, \bibinfo {author} {\bibfnamefont {J.}~\bibnamefont {Qian}}, \bibinfo {author} {\bibfnamefont {A.}~\bibnamefont {Clerk}}, \bibinfo {author} {\bibfnamefont {F.}~\bibnamefont {Marquardt}},\ and\ \bibinfo {author} {\bibfnamefont {K.}~\bibnamefont {Hammerer}},\ }\href@noop {} {\bibfield  {journal} {\bibinfo  {journal} {Physical Review X}\ }\textbf {\bibinfo {volume} {4}},\ \bibinfo {pages} {011015} (\bibinfo {year} {2014})}\BibitemShut {NoStop}%
\bibitem [{\citenamefont {Schulz}\ \emph {et~al.}(2016)\citenamefont {Schulz}, \citenamefont {Alvermann}, \citenamefont {Bakemeier},\ and\ \citenamefont {Fehske}}]{schulz2016optomechanical}%
  \BibitemOpen
  \bibfield  {author} {\bibinfo {author} {\bibfnamefont {C.}~\bibnamefont {Schulz}}, \bibinfo {author} {\bibfnamefont {A.}~\bibnamefont {Alvermann}}, \bibinfo {author} {\bibfnamefont {L.}~\bibnamefont {Bakemeier}},\ and\ \bibinfo {author} {\bibfnamefont {H.}~\bibnamefont {Fehske}},\ }\href@noop {} {\bibfield  {journal} {\bibinfo  {journal} {Europhysics Letters}\ }\textbf {\bibinfo {volume} {113}},\ \bibinfo {pages} {64002} (\bibinfo {year} {2016})}\BibitemShut {NoStop}%
\bibitem [{\citenamefont {Roque}\ \emph {et~al.}(2020)\citenamefont {Roque}, \citenamefont {Marquardt},\ and\ \citenamefont {Yevtushenko}}]{roque2020nonlinear}%
  \BibitemOpen
  \bibfield  {author} {\bibinfo {author} {\bibfnamefont {T.~F.}\ \bibnamefont {Roque}}, \bibinfo {author} {\bibfnamefont {F.}~\bibnamefont {Marquardt}},\ and\ \bibinfo {author} {\bibfnamefont {O.~M.}\ \bibnamefont {Yevtushenko}},\ }\href@noop {} {\bibfield  {journal} {\bibinfo  {journal} {New Journal of Physics}\ }\textbf {\bibinfo {volume} {22}},\ \bibinfo {pages} {013049} (\bibinfo {year} {2020})}\BibitemShut {NoStop}%
\bibitem [{\citenamefont {Bakemeier}\ \emph {et~al.}(2015)\citenamefont {Bakemeier}, \citenamefont {Alvermann},\ and\ \citenamefont {Fehske}}]{bakemeier2015route}%
  \BibitemOpen
  \bibfield  {author} {\bibinfo {author} {\bibfnamefont {L.}~\bibnamefont {Bakemeier}}, \bibinfo {author} {\bibfnamefont {A.}~\bibnamefont {Alvermann}},\ and\ \bibinfo {author} {\bibfnamefont {H.}~\bibnamefont {Fehske}},\ }\href@noop {} {\bibfield  {journal} {\bibinfo  {journal} {Physical review letters}\ }\textbf {\bibinfo {volume} {114}},\ \bibinfo {pages} {013601} (\bibinfo {year} {2015})}\BibitemShut {NoStop}%
\bibitem [{\citenamefont {L{\"u}}\ \emph {et~al.}(2015)\citenamefont {L{\"u}}, \citenamefont {Jing}, \citenamefont {Ma},\ and\ \citenamefont {Wu}}]{lu2015pt}%
  \BibitemOpen
  \bibfield  {author} {\bibinfo {author} {\bibfnamefont {X.-Y.}\ \bibnamefont {L{\"u}}}, \bibinfo {author} {\bibfnamefont {H.}~\bibnamefont {Jing}}, \bibinfo {author} {\bibfnamefont {J.-Y.}\ \bibnamefont {Ma}},\ and\ \bibinfo {author} {\bibfnamefont {Y.}~\bibnamefont {Wu}},\ }\href@noop {} {\bibfield  {journal} {\bibinfo  {journal} {Physical review letters}\ }\textbf {\bibinfo {volume} {114}},\ \bibinfo {pages} {253601} (\bibinfo {year} {2015})}\BibitemShut {NoStop}%
\bibitem [{\citenamefont {Kippenberg}\ \emph {et~al.}(2005)\citenamefont {Kippenberg}, \citenamefont {Rokhsari}, \citenamefont {Carmon}, \citenamefont {Scherer},\ and\ \citenamefont {Vahala}}]{kippenberg2005analysis}%
  \BibitemOpen
  \bibfield  {author} {\bibinfo {author} {\bibfnamefont {T.}~\bibnamefont {Kippenberg}}, \bibinfo {author} {\bibfnamefont {H.}~\bibnamefont {Rokhsari}}, \bibinfo {author} {\bibfnamefont {T.}~\bibnamefont {Carmon}}, \bibinfo {author} {\bibfnamefont {A.}~\bibnamefont {Scherer}},\ and\ \bibinfo {author} {\bibfnamefont {K.}~\bibnamefont {Vahala}},\ }\href@noop {} {\bibfield  {journal} {\bibinfo  {journal} {Physical Review Letters}\ }\textbf {\bibinfo {volume} {95}},\ \bibinfo {pages} {033901} (\bibinfo {year} {2005})}\BibitemShut {NoStop}%
\bibitem [{\citenamefont {Metzger}\ \emph {et~al.}(2008)\citenamefont {Metzger}, \citenamefont {Ludwig}, \citenamefont {Neuenhahn}, \citenamefont {Ortlieb}, \citenamefont {Favero}, \citenamefont {Karrai},\ and\ \citenamefont {Marquardt}}]{metzger2008self}%
  \BibitemOpen
  \bibfield  {author} {\bibinfo {author} {\bibfnamefont {C.}~\bibnamefont {Metzger}}, \bibinfo {author} {\bibfnamefont {M.}~\bibnamefont {Ludwig}}, \bibinfo {author} {\bibfnamefont {C.}~\bibnamefont {Neuenhahn}}, \bibinfo {author} {\bibfnamefont {A.}~\bibnamefont {Ortlieb}}, \bibinfo {author} {\bibfnamefont {I.}~\bibnamefont {Favero}}, \bibinfo {author} {\bibfnamefont {.~f.~K.}\ \bibnamefont {Karrai}},\ and\ \bibinfo {author} {\bibfnamefont {F.}~\bibnamefont {Marquardt}},\ }\href@noop {} {\bibfield  {journal} {\bibinfo  {journal} {Physical review letters}\ }\textbf {\bibinfo {volume} {101}},\ \bibinfo {pages} {133903} (\bibinfo {year} {2008})}\BibitemShut {NoStop}%
\bibitem [{\citenamefont {Carmon}\ \emph {et~al.}(2007)\citenamefont {Carmon}, \citenamefont {Cross},\ and\ \citenamefont {Vahala}}]{carmon2007chaotic}%
  \BibitemOpen
  \bibfield  {author} {\bibinfo {author} {\bibfnamefont {T.}~\bibnamefont {Carmon}}, \bibinfo {author} {\bibfnamefont {M.}~\bibnamefont {Cross}},\ and\ \bibinfo {author} {\bibfnamefont {K.~J.}\ \bibnamefont {Vahala}},\ }\href@noop {} {\bibfield  {journal} {\bibinfo  {journal} {Physical review letters}\ }\textbf {\bibinfo {volume} {98}},\ \bibinfo {pages} {167203} (\bibinfo {year} {2007})}\BibitemShut {NoStop}%
\bibitem [{\citenamefont {Navarro-Urrios}\ \emph {et~al.}(2017)\citenamefont {Navarro-Urrios}, \citenamefont {Capuj}, \citenamefont {Colombano}, \citenamefont {Garc{\'\i}a}, \citenamefont {Sledzinska}, \citenamefont {Alzina}, \citenamefont {Griol}, \citenamefont {Mart{\'\i}nez},\ and\ \citenamefont {Sotomayor-Torres}}]{navarro2017nonlinear}%
  \BibitemOpen
  \bibfield  {author} {\bibinfo {author} {\bibfnamefont {D.}~\bibnamefont {Navarro-Urrios}}, \bibinfo {author} {\bibfnamefont {N.~E.}\ \bibnamefont {Capuj}}, \bibinfo {author} {\bibfnamefont {M.~F.}\ \bibnamefont {Colombano}}, \bibinfo {author} {\bibfnamefont {P.~D.}\ \bibnamefont {Garc{\'\i}a}}, \bibinfo {author} {\bibfnamefont {M.}~\bibnamefont {Sledzinska}}, \bibinfo {author} {\bibfnamefont {F.}~\bibnamefont {Alzina}}, \bibinfo {author} {\bibfnamefont {A.}~\bibnamefont {Griol}}, \bibinfo {author} {\bibfnamefont {A.}~\bibnamefont {Mart{\'\i}nez}},\ and\ \bibinfo {author} {\bibfnamefont {C.~M.}\ \bibnamefont {Sotomayor-Torres}},\ }\href@noop {} {\bibfield  {journal} {\bibinfo  {journal} {Nature communications}\ }\textbf {\bibinfo {volume} {8}},\ \bibinfo {pages} {14965} (\bibinfo {year} {2017})}\BibitemShut {NoStop}%
\bibitem [{\citenamefont {Das}\ \emph {et~al.}(2023)\citenamefont {Das}, \citenamefont {Majumder}, \citenamefont {Sahu}, \citenamefont {Singhal}, \citenamefont {Bera},\ and\ \citenamefont {Singh}}]{das2023instabilities}%
  \BibitemOpen
  \bibfield  {author} {\bibinfo {author} {\bibfnamefont {S.~R.}\ \bibnamefont {Das}}, \bibinfo {author} {\bibfnamefont {S.}~\bibnamefont {Majumder}}, \bibinfo {author} {\bibfnamefont {S.~K.}\ \bibnamefont {Sahu}}, \bibinfo {author} {\bibfnamefont {U.}~\bibnamefont {Singhal}}, \bibinfo {author} {\bibfnamefont {T.}~\bibnamefont {Bera}},\ and\ \bibinfo {author} {\bibfnamefont {V.}~\bibnamefont {Singh}},\ }\href@noop {} {\bibfield  {journal} {\bibinfo  {journal} {Physical Review Letters}\ }\textbf {\bibinfo {volume} {131}},\ \bibinfo {pages} {067001} (\bibinfo {year} {2023})}\BibitemShut {NoStop}%
\bibitem [{\citenamefont {Krause}\ \emph {et~al.}(2015)\citenamefont {Krause}, \citenamefont {Hill}, \citenamefont {Ludwig}, \citenamefont {Safavi-Naeini}, \citenamefont {Chan}, \citenamefont {Marquardt},\ and\ \citenamefont {Painter}}]{krause2015nonlinear}%
  \BibitemOpen
  \bibfield  {author} {\bibinfo {author} {\bibfnamefont {A.~G.}\ \bibnamefont {Krause}}, \bibinfo {author} {\bibfnamefont {J.~T.}\ \bibnamefont {Hill}}, \bibinfo {author} {\bibfnamefont {M.}~\bibnamefont {Ludwig}}, \bibinfo {author} {\bibfnamefont {A.~H.}\ \bibnamefont {Safavi-Naeini}}, \bibinfo {author} {\bibfnamefont {J.}~\bibnamefont {Chan}}, \bibinfo {author} {\bibfnamefont {F.}~\bibnamefont {Marquardt}},\ and\ \bibinfo {author} {\bibfnamefont {O.}~\bibnamefont {Painter}},\ }\href@noop {} {\bibfield  {journal} {\bibinfo  {journal} {Physical Review Letters}\ }\textbf {\bibinfo {volume} {115}},\ \bibinfo {pages} {233601} (\bibinfo {year} {2015})}\BibitemShut {NoStop}%
\bibitem [{\citenamefont {Buters}\ \emph {et~al.}(2015)\citenamefont {Buters}, \citenamefont {Eerkens}, \citenamefont {Heeck}, \citenamefont {Weaver}, \citenamefont {Pepper}, \citenamefont {de~Man},\ and\ \citenamefont {Bouwmeester}}]{buters2015experimental}%
  \BibitemOpen
  \bibfield  {author} {\bibinfo {author} {\bibfnamefont {F.~M.}\ \bibnamefont {Buters}}, \bibinfo {author} {\bibfnamefont {H.~J.}\ \bibnamefont {Eerkens}}, \bibinfo {author} {\bibfnamefont {K.}~\bibnamefont {Heeck}}, \bibinfo {author} {\bibfnamefont {M.~J.}\ \bibnamefont {Weaver}}, \bibinfo {author} {\bibfnamefont {B.}~\bibnamefont {Pepper}}, \bibinfo {author} {\bibfnamefont {S.}~\bibnamefont {de~Man}},\ and\ \bibinfo {author} {\bibfnamefont {D.}~\bibnamefont {Bouwmeester}},\ }\href@noop {} {\bibfield  {journal} {\bibinfo  {journal} {Physical Review A}\ }\textbf {\bibinfo {volume} {92}},\ \bibinfo {pages} {013811} (\bibinfo {year} {2015})}\BibitemShut {NoStop}%
\bibitem [{\citenamefont {Heinrich}\ \emph {et~al.}(2011)\citenamefont {Heinrich}, \citenamefont {Ludwig}, \citenamefont {Qian}, \citenamefont {Kubala},\ and\ \citenamefont {Marquardt}}]{heinrich2011collective}%
  \BibitemOpen
  \bibfield  {author} {\bibinfo {author} {\bibfnamefont {G.}~\bibnamefont {Heinrich}}, \bibinfo {author} {\bibfnamefont {M.}~\bibnamefont {Ludwig}}, \bibinfo {author} {\bibfnamefont {J.}~\bibnamefont {Qian}}, \bibinfo {author} {\bibfnamefont {B.}~\bibnamefont {Kubala}},\ and\ \bibinfo {author} {\bibfnamefont {F.}~\bibnamefont {Marquardt}},\ }\href@noop {} {\bibfield  {journal} {\bibinfo  {journal} {Physical review letters}\ }\textbf {\bibinfo {volume} {107}},\ \bibinfo {pages} {043603} (\bibinfo {year} {2011})}\BibitemShut {NoStop}%
\bibitem [{\citenamefont {Zhang}\ \emph {et~al.}(2012)\citenamefont {Zhang}, \citenamefont {Wiederhecker}, \citenamefont {Manipatruni}, \citenamefont {Barnard}, \citenamefont {McEuen},\ and\ \citenamefont {Lipson}}]{zhang2012synchronization}%
  \BibitemOpen
  \bibfield  {author} {\bibinfo {author} {\bibfnamefont {M.}~\bibnamefont {Zhang}}, \bibinfo {author} {\bibfnamefont {G.~S.}\ \bibnamefont {Wiederhecker}}, \bibinfo {author} {\bibfnamefont {S.}~\bibnamefont {Manipatruni}}, \bibinfo {author} {\bibfnamefont {A.}~\bibnamefont {Barnard}}, \bibinfo {author} {\bibfnamefont {P.}~\bibnamefont {McEuen}},\ and\ \bibinfo {author} {\bibfnamefont {M.}~\bibnamefont {Lipson}},\ }\href@noop {} {\bibfield  {journal} {\bibinfo  {journal} {Physical review letters}\ }\textbf {\bibinfo {volume} {109}},\ \bibinfo {pages} {233906} (\bibinfo {year} {2012})}\BibitemShut {NoStop}%
\bibitem [{\citenamefont {Weiss}\ \emph {et~al.}(2016)\citenamefont {Weiss}, \citenamefont {Kronwald},\ and\ \citenamefont {Marquardt}}]{weiss2016noise}%
  \BibitemOpen
  \bibfield  {author} {\bibinfo {author} {\bibfnamefont {T.}~\bibnamefont {Weiss}}, \bibinfo {author} {\bibfnamefont {A.}~\bibnamefont {Kronwald}},\ and\ \bibinfo {author} {\bibfnamefont {F.}~\bibnamefont {Marquardt}},\ }\href@noop {} {\bibfield  {journal} {\bibinfo  {journal} {New Journal of Physics}\ }\textbf {\bibinfo {volume} {18}},\ \bibinfo {pages} {013043} (\bibinfo {year} {2016})}\BibitemShut {NoStop}%
\bibitem [{\citenamefont {Ghobadi}\ \emph {et~al.}(2011)\citenamefont {Ghobadi}, \citenamefont {Bahrampour},\ and\ \citenamefont {Simon}}]{ghobadi2011quantum}%
  \BibitemOpen
  \bibfield  {author} {\bibinfo {author} {\bibfnamefont {R.}~\bibnamefont {Ghobadi}}, \bibinfo {author} {\bibfnamefont {A.}~\bibnamefont {Bahrampour}},\ and\ \bibinfo {author} {\bibfnamefont {C.}~\bibnamefont {Simon}},\ }\href@noop {} {\bibfield  {journal} {\bibinfo  {journal} {Physical Review A—Atomic, Molecular, and Optical Physics}\ }\textbf {\bibinfo {volume} {84}},\ \bibinfo {pages} {033846} (\bibinfo {year} {2011})}\BibitemShut {NoStop}%
\bibitem [{\citenamefont {Bibak}\ \emph {et~al.}(2023)\citenamefont {Bibak}, \citenamefont {Deli{\'c}}, \citenamefont {Aspelmeyer},\ and\ \citenamefont {Daki{\'c}}}]{bibak2023dissipative}%
  \BibitemOpen
  \bibfield  {author} {\bibinfo {author} {\bibfnamefont {F.}~\bibnamefont {Bibak}}, \bibinfo {author} {\bibfnamefont {U.}~\bibnamefont {Deli{\'c}}}, \bibinfo {author} {\bibfnamefont {M.}~\bibnamefont {Aspelmeyer}},\ and\ \bibinfo {author} {\bibfnamefont {B.}~\bibnamefont {Daki{\'c}}},\ }\href@noop {} {\bibfield  {journal} {\bibinfo  {journal} {Physical Review A}\ }\textbf {\bibinfo {volume} {107}},\ \bibinfo {pages} {053505} (\bibinfo {year} {2023})}\BibitemShut {NoStop}%
\bibitem [{\citenamefont {Wang}\ \emph {et~al.}(2024)\citenamefont {Wang}, \citenamefont {Nori},\ and\ \citenamefont {Xiang}}]{wang2024quantum}%
  \BibitemOpen
  \bibfield  {author} {\bibinfo {author} {\bibfnamefont {B.}~\bibnamefont {Wang}}, \bibinfo {author} {\bibfnamefont {F.}~\bibnamefont {Nori}},\ and\ \bibinfo {author} {\bibfnamefont {Z.-L.}\ \bibnamefont {Xiang}},\ }\href@noop {} {\bibfield  {journal} {\bibinfo  {journal} {Physical Review Letters}\ }\textbf {\bibinfo {volume} {132}},\ \bibinfo {pages} {053601} (\bibinfo {year} {2024})}\BibitemShut {NoStop}%
\bibitem [{\citenamefont {Rameshti}\ \emph {et~al.}(2022)\citenamefont {Rameshti}, \citenamefont {Kusminskiy}, \citenamefont {Haigh}, \citenamefont {Usami}, \citenamefont {Lachance-Quirion}, \citenamefont {Nakamura}, \citenamefont {Hu}, \citenamefont {Tang}, \citenamefont {Bauer},\ and\ \citenamefont {Blanter}}]{rameshti2022cavity}%
  \BibitemOpen
  \bibfield  {author} {\bibinfo {author} {\bibfnamefont {B.~Z.}\ \bibnamefont {Rameshti}}, \bibinfo {author} {\bibfnamefont {S.~V.}\ \bibnamefont {Kusminskiy}}, \bibinfo {author} {\bibfnamefont {J.~A.}\ \bibnamefont {Haigh}}, \bibinfo {author} {\bibfnamefont {K.}~\bibnamefont {Usami}}, \bibinfo {author} {\bibfnamefont {D.}~\bibnamefont {Lachance-Quirion}}, \bibinfo {author} {\bibfnamefont {Y.}~\bibnamefont {Nakamura}}, \bibinfo {author} {\bibfnamefont {C.-M.}\ \bibnamefont {Hu}}, \bibinfo {author} {\bibfnamefont {H.~X.}\ \bibnamefont {Tang}}, \bibinfo {author} {\bibfnamefont {G.~E.}\ \bibnamefont {Bauer}},\ and\ \bibinfo {author} {\bibfnamefont {Y.~M.}\ \bibnamefont {Blanter}},\ }\href@noop {} {\bibfield  {journal} {\bibinfo  {journal} {Physics Reports}\ }\textbf {\bibinfo {volume} {979}},\ \bibinfo {pages} {1} (\bibinfo {year} {2022})}\BibitemShut {NoStop}%
\bibitem [{\citenamefont {Yuan}\ \emph {et~al.}(2022)\citenamefont {Yuan}, \citenamefont {Cao}, \citenamefont {Kamra}, \citenamefont {Duine},\ and\ \citenamefont {Yan}}]{yuan2022quantum}%
  \BibitemOpen
  \bibfield  {author} {\bibinfo {author} {\bibfnamefont {H.}~\bibnamefont {Yuan}}, \bibinfo {author} {\bibfnamefont {Y.}~\bibnamefont {Cao}}, \bibinfo {author} {\bibfnamefont {A.}~\bibnamefont {Kamra}}, \bibinfo {author} {\bibfnamefont {R.~A.}\ \bibnamefont {Duine}},\ and\ \bibinfo {author} {\bibfnamefont {P.}~\bibnamefont {Yan}},\ }\href@noop {} {\bibfield  {journal} {\bibinfo  {journal} {Physics Reports}\ }\textbf {\bibinfo {volume} {965}},\ \bibinfo {pages} {1} (\bibinfo {year} {2022})}\BibitemShut {NoStop}%
\bibitem [{\citenamefont {Huebl}\ \emph {et~al.}(2013)\citenamefont {Huebl}, \citenamefont {Zollitsch}, \citenamefont {Lotze}, \citenamefont {Hocke}, \citenamefont {Greifenstein}, \citenamefont {Marx}, \citenamefont {Gross},\ and\ \citenamefont {Goennenwein}}]{huebl2013high}%
  \BibitemOpen
  \bibfield  {author} {\bibinfo {author} {\bibfnamefont {H.}~\bibnamefont {Huebl}}, \bibinfo {author} {\bibfnamefont {C.~W.}\ \bibnamefont {Zollitsch}}, \bibinfo {author} {\bibfnamefont {J.}~\bibnamefont {Lotze}}, \bibinfo {author} {\bibfnamefont {F.}~\bibnamefont {Hocke}}, \bibinfo {author} {\bibfnamefont {M.}~\bibnamefont {Greifenstein}}, \bibinfo {author} {\bibfnamefont {A.}~\bibnamefont {Marx}}, \bibinfo {author} {\bibfnamefont {R.}~\bibnamefont {Gross}},\ and\ \bibinfo {author} {\bibfnamefont {S.~T.}\ \bibnamefont {Goennenwein}},\ }\href@noop {} {\bibfield  {journal} {\bibinfo  {journal} {Physical Review Letters}\ }\textbf {\bibinfo {volume} {111}},\ \bibinfo {pages} {127003} (\bibinfo {year} {2013})}\BibitemShut {NoStop}%
\bibitem [{\citenamefont {Tabuchi}\ \emph {et~al.}(2014)\citenamefont {Tabuchi}, \citenamefont {Ishino}, \citenamefont {Ishikawa}, \citenamefont {Yamazaki}, \citenamefont {Usami},\ and\ \citenamefont {Nakamura}}]{tabuchi2014hybridizing}%
  \BibitemOpen
  \bibfield  {author} {\bibinfo {author} {\bibfnamefont {Y.}~\bibnamefont {Tabuchi}}, \bibinfo {author} {\bibfnamefont {S.}~\bibnamefont {Ishino}}, \bibinfo {author} {\bibfnamefont {T.}~\bibnamefont {Ishikawa}}, \bibinfo {author} {\bibfnamefont {R.}~\bibnamefont {Yamazaki}}, \bibinfo {author} {\bibfnamefont {K.}~\bibnamefont {Usami}},\ and\ \bibinfo {author} {\bibfnamefont {Y.}~\bibnamefont {Nakamura}},\ }\href@noop {} {\bibfield  {journal} {\bibinfo  {journal} {Physical review letters}\ }\textbf {\bibinfo {volume} {113}},\ \bibinfo {pages} {083603} (\bibinfo {year} {2014})}\BibitemShut {NoStop}%
\bibitem [{\citenamefont {Zhang}\ \emph {et~al.}(2014)\citenamefont {Zhang}, \citenamefont {Zou}, \citenamefont {Jiang},\ and\ \citenamefont {Tang}}]{zhang2014strongly}%
  \BibitemOpen
  \bibfield  {author} {\bibinfo {author} {\bibfnamefont {X.}~\bibnamefont {Zhang}}, \bibinfo {author} {\bibfnamefont {C.-L.}\ \bibnamefont {Zou}}, \bibinfo {author} {\bibfnamefont {L.}~\bibnamefont {Jiang}},\ and\ \bibinfo {author} {\bibfnamefont {H.~X.}\ \bibnamefont {Tang}},\ }\href@noop {} {\bibfield  {journal} {\bibinfo  {journal} {Physical review letters}\ }\textbf {\bibinfo {volume} {113}},\ \bibinfo {pages} {156401} (\bibinfo {year} {2014})}\BibitemShut {NoStop}%
\bibitem [{\citenamefont {Bai}\ \emph {et~al.}(2015)\citenamefont {Bai}, \citenamefont {Harder}, \citenamefont {Chen}, \citenamefont {Fan}, \citenamefont {Xiao},\ and\ \citenamefont {Hu}}]{bai2015spin}%
  \BibitemOpen
  \bibfield  {author} {\bibinfo {author} {\bibfnamefont {L.}~\bibnamefont {Bai}}, \bibinfo {author} {\bibfnamefont {M.}~\bibnamefont {Harder}}, \bibinfo {author} {\bibfnamefont {Y.}~\bibnamefont {Chen}}, \bibinfo {author} {\bibfnamefont {X.}~\bibnamefont {Fan}}, \bibinfo {author} {\bibfnamefont {J.}~\bibnamefont {Xiao}},\ and\ \bibinfo {author} {\bibfnamefont {C.-M.}\ \bibnamefont {Hu}},\ }\href@noop {} {\bibfield  {journal} {\bibinfo  {journal} {Physical Review Letters}\ }\textbf {\bibinfo {volume} {114}},\ \bibinfo {pages} {227201} (\bibinfo {year} {2015})}\BibitemShut {NoStop}%
\bibitem [{\citenamefont {Osada}\ \emph {et~al.}(2016)\citenamefont {Osada}, \citenamefont {Hisatomi}, \citenamefont {Noguchi}, \citenamefont {Tabuchi}, \citenamefont {Yamazaki}, \citenamefont {Usami}, \citenamefont {Sadgrove}, \citenamefont {Yalla}, \citenamefont {Nomura},\ and\ \citenamefont {Nakamura}}]{osada2016cavity}%
  \BibitemOpen
  \bibfield  {author} {\bibinfo {author} {\bibfnamefont {A.}~\bibnamefont {Osada}}, \bibinfo {author} {\bibfnamefont {R.}~\bibnamefont {Hisatomi}}, \bibinfo {author} {\bibfnamefont {A.}~\bibnamefont {Noguchi}}, \bibinfo {author} {\bibfnamefont {Y.}~\bibnamefont {Tabuchi}}, \bibinfo {author} {\bibfnamefont {R.}~\bibnamefont {Yamazaki}}, \bibinfo {author} {\bibfnamefont {K.}~\bibnamefont {Usami}}, \bibinfo {author} {\bibfnamefont {M.}~\bibnamefont {Sadgrove}}, \bibinfo {author} {\bibfnamefont {R.}~\bibnamefont {Yalla}}, \bibinfo {author} {\bibfnamefont {M.}~\bibnamefont {Nomura}},\ and\ \bibinfo {author} {\bibfnamefont {Y.}~\bibnamefont {Nakamura}},\ }\href@noop {} {\bibfield  {journal} {\bibinfo  {journal} {Physical review letters}\ }\textbf {\bibinfo {volume} {116}},\ \bibinfo {pages} {223601} (\bibinfo {year} {2016})}\BibitemShut {NoStop}%
\bibitem [{\citenamefont {Hisatomi}\ \emph {et~al.}(2016)\citenamefont {Hisatomi}, \citenamefont {Osada}, \citenamefont {Tabuchi}, \citenamefont {Ishikawa}, \citenamefont {Noguchi}, \citenamefont {Yamazaki}, \citenamefont {Usami},\ and\ \citenamefont {Nakamura}}]{hisatomi2016bidirectional}%
  \BibitemOpen
  \bibfield  {author} {\bibinfo {author} {\bibfnamefont {R.}~\bibnamefont {Hisatomi}}, \bibinfo {author} {\bibfnamefont {A.}~\bibnamefont {Osada}}, \bibinfo {author} {\bibfnamefont {Y.}~\bibnamefont {Tabuchi}}, \bibinfo {author} {\bibfnamefont {T.}~\bibnamefont {Ishikawa}}, \bibinfo {author} {\bibfnamefont {A.}~\bibnamefont {Noguchi}}, \bibinfo {author} {\bibfnamefont {R.}~\bibnamefont {Yamazaki}}, \bibinfo {author} {\bibfnamefont {K.}~\bibnamefont {Usami}},\ and\ \bibinfo {author} {\bibfnamefont {Y.}~\bibnamefont {Nakamura}},\ }\href@noop {} {\bibfield  {journal} {\bibinfo  {journal} {Physical Review B}\ }\textbf {\bibinfo {volume} {93}},\ \bibinfo {pages} {174427} (\bibinfo {year} {2016})}\BibitemShut {NoStop}%
\bibitem [{\citenamefont {Zhang}\ \emph {et~al.}(2015)\citenamefont {Zhang}, \citenamefont {Zou}, \citenamefont {Zhu}, \citenamefont {Marquardt}, \citenamefont {Jiang},\ and\ \citenamefont {Tang}}]{zhang2015magnon}%
  \BibitemOpen
  \bibfield  {author} {\bibinfo {author} {\bibfnamefont {X.}~\bibnamefont {Zhang}}, \bibinfo {author} {\bibfnamefont {C.-L.}\ \bibnamefont {Zou}}, \bibinfo {author} {\bibfnamefont {N.}~\bibnamefont {Zhu}}, \bibinfo {author} {\bibfnamefont {F.}~\bibnamefont {Marquardt}}, \bibinfo {author} {\bibfnamefont {L.}~\bibnamefont {Jiang}},\ and\ \bibinfo {author} {\bibfnamefont {H.~X.}\ \bibnamefont {Tang}},\ }\href@noop {} {\bibfield  {journal} {\bibinfo  {journal} {Nature communications}\ }\textbf {\bibinfo {volume} {6}},\ \bibinfo {pages} {8914} (\bibinfo {year} {2015})}\BibitemShut {NoStop}%
\bibitem [{\citenamefont {Zhang}\ \emph {et~al.}(2016)\citenamefont {Zhang}, \citenamefont {Zou}, \citenamefont {Jiang},\ and\ \citenamefont {Tang}}]{zhang2016cavity}%
  \BibitemOpen
  \bibfield  {author} {\bibinfo {author} {\bibfnamefont {X.}~\bibnamefont {Zhang}}, \bibinfo {author} {\bibfnamefont {C.-L.}\ \bibnamefont {Zou}}, \bibinfo {author} {\bibfnamefont {L.}~\bibnamefont {Jiang}},\ and\ \bibinfo {author} {\bibfnamefont {H.~X.}\ \bibnamefont {Tang}},\ }\href@noop {} {\bibfield  {journal} {\bibinfo  {journal} {Science advances}\ }\textbf {\bibinfo {volume} {2}},\ \bibinfo {pages} {e1501286} (\bibinfo {year} {2016})}\BibitemShut {NoStop}%
\bibitem [{\citenamefont {Potts}\ \emph {et~al.}(2021)\citenamefont {Potts}, \citenamefont {Varga}, \citenamefont {Bittencourt}, \citenamefont {Kusminskiy},\ and\ \citenamefont {Davis}}]{potts2021dynamical}%
  \BibitemOpen
  \bibfield  {author} {\bibinfo {author} {\bibfnamefont {C.~A.}\ \bibnamefont {Potts}}, \bibinfo {author} {\bibfnamefont {E.}~\bibnamefont {Varga}}, \bibinfo {author} {\bibfnamefont {V.~A.}\ \bibnamefont {Bittencourt}}, \bibinfo {author} {\bibfnamefont {S.~V.}\ \bibnamefont {Kusminskiy}},\ and\ \bibinfo {author} {\bibfnamefont {J.~P.}\ \bibnamefont {Davis}},\ }\href@noop {} {\bibfield  {journal} {\bibinfo  {journal} {Physical Review X}\ }\textbf {\bibinfo {volume} {11}},\ \bibinfo {pages} {031053} (\bibinfo {year} {2021})}\BibitemShut {NoStop}%
\bibitem [{\citenamefont {Li}\ \emph {et~al.}(2019)\citenamefont {Li}, \citenamefont {Zhu},\ and\ \citenamefont {Agarwal}}]{li2019squeezed}%
  \BibitemOpen
  \bibfield  {author} {\bibinfo {author} {\bibfnamefont {J.}~\bibnamefont {Li}}, \bibinfo {author} {\bibfnamefont {S.-Y.}\ \bibnamefont {Zhu}},\ and\ \bibinfo {author} {\bibfnamefont {G.}~\bibnamefont {Agarwal}},\ }\href@noop {} {\bibfield  {journal} {\bibinfo  {journal} {Physical Review A}\ }\textbf {\bibinfo {volume} {99}},\ \bibinfo {pages} {021801} (\bibinfo {year} {2019})}\BibitemShut {NoStop}%
\bibitem [{\citenamefont {Li}\ \emph {et~al.}(2018)\citenamefont {Li}, \citenamefont {Zhu},\ and\ \citenamefont {Agarwal}}]{li2018magnon}%
  \BibitemOpen
  \bibfield  {author} {\bibinfo {author} {\bibfnamefont {J.}~\bibnamefont {Li}}, \bibinfo {author} {\bibfnamefont {S.-Y.}\ \bibnamefont {Zhu}},\ and\ \bibinfo {author} {\bibfnamefont {G.}~\bibnamefont {Agarwal}},\ }\href@noop {} {\bibfield  {journal} {\bibinfo  {journal} {Physical review letters}\ }\textbf {\bibinfo {volume} {121}},\ \bibinfo {pages} {203601} (\bibinfo {year} {2018})}\BibitemShut {NoStop}%
\bibitem [{\citenamefont {Ding}\ \emph {et~al.}(2019)\citenamefont {Ding}, \citenamefont {Zheng},\ and\ \citenamefont {Li}}]{ding2019phonon}%
  \BibitemOpen
  \bibfield  {author} {\bibinfo {author} {\bibfnamefont {M.-S.}\ \bibnamefont {Ding}}, \bibinfo {author} {\bibfnamefont {L.}~\bibnamefont {Zheng}},\ and\ \bibinfo {author} {\bibfnamefont {C.}~\bibnamefont {Li}},\ }\href@noop {} {\bibfield  {journal} {\bibinfo  {journal} {Scientific reports}\ }\textbf {\bibinfo {volume} {9}},\ \bibinfo {pages} {15723} (\bibinfo {year} {2019})}\BibitemShut {NoStop}%
\bibitem [{\citenamefont {Peng}\ \emph {et~al.}(2024)\citenamefont {Peng}, \citenamefont {Liu}, \citenamefont {Yu},\ and\ \citenamefont {Xiong}}]{peng2024ultra}%
  \BibitemOpen
  \bibfield  {author} {\bibinfo {author} {\bibfnamefont {J.}~\bibnamefont {Peng}}, \bibinfo {author} {\bibfnamefont {Z.-X.}\ \bibnamefont {Liu}}, \bibinfo {author} {\bibfnamefont {Y.-F.}\ \bibnamefont {Yu}},\ and\ \bibinfo {author} {\bibfnamefont {H.}~\bibnamefont {Xiong}},\ }\href@noop {} {\bibfield  {journal} {\bibinfo  {journal} {arXiv preprint arXiv:2407.13145}\ } (\bibinfo {year} {2024})}\BibitemShut {NoStop}%
\bibitem [{\citenamefont {Shen}\ \emph {et~al.}(2022)\citenamefont {Shen}, \citenamefont {Li}, \citenamefont {Fan}, \citenamefont {Wang},\ and\ \citenamefont {You}}]{shen2022mechanical}%
  \BibitemOpen
  \bibfield  {author} {\bibinfo {author} {\bibfnamefont {R.-C.}\ \bibnamefont {Shen}}, \bibinfo {author} {\bibfnamefont {J.}~\bibnamefont {Li}}, \bibinfo {author} {\bibfnamefont {Z.-Y.}\ \bibnamefont {Fan}}, \bibinfo {author} {\bibfnamefont {Y.-P.}\ \bibnamefont {Wang}},\ and\ \bibinfo {author} {\bibfnamefont {J.}~\bibnamefont {You}},\ }\href@noop {} {\bibfield  {journal} {\bibinfo  {journal} {Physical Review Letters}\ }\textbf {\bibinfo {volume} {129}},\ \bibinfo {pages} {123601} (\bibinfo {year} {2022})}\BibitemShut {NoStop}%
\bibitem [{\citenamefont {Yang}\ \emph {et~al.}(2021)\citenamefont {Yang}, \citenamefont {Jin}, \citenamefont {Jin}, \citenamefont {Liu}, \citenamefont {Liu},\ and\ \citenamefont {Yang}}]{yang2021bistability}%
  \BibitemOpen
  \bibfield  {author} {\bibinfo {author} {\bibfnamefont {Z.-B.}\ \bibnamefont {Yang}}, \bibinfo {author} {\bibfnamefont {H.}~\bibnamefont {Jin}}, \bibinfo {author} {\bibfnamefont {J.-W.}\ \bibnamefont {Jin}}, \bibinfo {author} {\bibfnamefont {J.-Y.}\ \bibnamefont {Liu}}, \bibinfo {author} {\bibfnamefont {H.-Y.}\ \bibnamefont {Liu}},\ and\ \bibinfo {author} {\bibfnamefont {R.-C.}\ \bibnamefont {Yang}},\ }\href@noop {} {\bibfield  {journal} {\bibinfo  {journal} {Physical Review Research}\ }\textbf {\bibinfo {volume} {3}},\ \bibinfo {pages} {023126} (\bibinfo {year} {2021})}\BibitemShut {NoStop}%
\bibitem [{\citenamefont {Sarma}\ \emph {et~al.}(2021)\citenamefont {Sarma}, \citenamefont {Chakraborty},\ and\ \citenamefont {Kalita}}]{sarma2021continuous}%
  \BibitemOpen
  \bibfield  {author} {\bibinfo {author} {\bibfnamefont {A.~K.}\ \bibnamefont {Sarma}}, \bibinfo {author} {\bibfnamefont {S.}~\bibnamefont {Chakraborty}},\ and\ \bibinfo {author} {\bibfnamefont {S.}~\bibnamefont {Kalita}},\ }\href@noop {} {\bibfield  {journal} {\bibinfo  {journal} {AVS Quantum Science}\ }\textbf {\bibinfo {volume} {3}} (\bibinfo {year} {2021})}\BibitemShut {NoStop}%
\bibitem [{\citenamefont {Adesso}\ \emph {et~al.}(2014)\citenamefont {Adesso}, \citenamefont {Ragy},\ and\ \citenamefont {Lee}}]{adesso2014continuous}%
  \BibitemOpen
  \bibfield  {author} {\bibinfo {author} {\bibfnamefont {G.}~\bibnamefont {Adesso}}, \bibinfo {author} {\bibfnamefont {S.}~\bibnamefont {Ragy}},\ and\ \bibinfo {author} {\bibfnamefont {A.~R.}\ \bibnamefont {Lee}},\ }\href@noop {} {\bibfield  {journal} {\bibinfo  {journal} {Open Systems \& Information Dynamics}\ }\textbf {\bibinfo {volume} {21}},\ \bibinfo {pages} {1440001} (\bibinfo {year} {2014})}\BibitemShut {NoStop}%
\bibitem [{\citenamefont {Vidal}\ and\ \citenamefont {Werner}(2002)}]{vidal2002computable}%
  \BibitemOpen
  \bibfield  {author} {\bibinfo {author} {\bibfnamefont {G.}~\bibnamefont {Vidal}}\ and\ \bibinfo {author} {\bibfnamefont {R.~F.}\ \bibnamefont {Werner}},\ }\href@noop {} {\bibfield  {journal} {\bibinfo  {journal} {Physical Review A}\ }\textbf {\bibinfo {volume} {65}},\ \bibinfo {pages} {032314} (\bibinfo {year} {2002})}\BibitemShut {NoStop}%
\bibitem [{\citenamefont {Yu}\ \emph {et~al.}(2020)\citenamefont {Yu}, \citenamefont {Shen},\ and\ \citenamefont {Li}}]{yu2020magnetostrictively}%
  \BibitemOpen
  \bibfield  {author} {\bibinfo {author} {\bibfnamefont {M.}~\bibnamefont {Yu}}, \bibinfo {author} {\bibfnamefont {H.}~\bibnamefont {Shen}},\ and\ \bibinfo {author} {\bibfnamefont {J.}~\bibnamefont {Li}},\ }\href@noop {} {\bibfield  {journal} {\bibinfo  {journal} {Physical Review Letters}\ }\textbf {\bibinfo {volume} {124}},\ \bibinfo {pages} {213604} (\bibinfo {year} {2020})}\BibitemShut {NoStop}%
\bibitem [{\citenamefont {Li}\ and\ \citenamefont {Zhu}(2019)}]{li2019entangling}%
  \BibitemOpen
  \bibfield  {author} {\bibinfo {author} {\bibfnamefont {J.}~\bibnamefont {Li}}\ and\ \bibinfo {author} {\bibfnamefont {S.-Y.}\ \bibnamefont {Zhu}},\ }\href@noop {} {\bibfield  {journal} {\bibinfo  {journal} {New Journal of Physics}\ }\textbf {\bibinfo {volume} {21}},\ \bibinfo {pages} {085001} (\bibinfo {year} {2019})}\BibitemShut {NoStop}%
\bibitem [{\citenamefont {Li}\ and\ \citenamefont {Gr{\"o}blacher}(2021)}]{li2021entangling}%
  \BibitemOpen
  \bibfield  {author} {\bibinfo {author} {\bibfnamefont {J.}~\bibnamefont {Li}}\ and\ \bibinfo {author} {\bibfnamefont {S.}~\bibnamefont {Gr{\"o}blacher}},\ }\href@noop {} {\bibfield  {journal} {\bibinfo  {journal} {Quantum Science and Technology}\ }\textbf {\bibinfo {volume} {6}},\ \bibinfo {pages} {024005} (\bibinfo {year} {2021})}\BibitemShut {NoStop}%
\bibitem [{\citenamefont {Meng}\ \emph {et~al.}(2020)\citenamefont {Meng}, \citenamefont {Deng}, \citenamefont {Zhu},\ and\ \citenamefont {Huang}}]{meng2020quantum}%
  \BibitemOpen
  \bibfield  {author} {\bibinfo {author} {\bibfnamefont {Q.~X.}\ \bibnamefont {Meng}}, \bibinfo {author} {\bibfnamefont {Z.~J.}\ \bibnamefont {Deng}}, \bibinfo {author} {\bibfnamefont {Z.}~\bibnamefont {Zhu}},\ and\ \bibinfo {author} {\bibfnamefont {L.}~\bibnamefont {Huang}},\ }\href@noop {} {\bibfield  {journal} {\bibinfo  {journal} {Physical Review A}\ }\textbf {\bibinfo {volume} {101}},\ \bibinfo {pages} {023838} (\bibinfo {year} {2020})}\BibitemShut {NoStop}%
\bibitem [{\citenamefont {Amitai}\ \emph {et~al.}(2017)\citenamefont {Amitai}, \citenamefont {L{\"o}rch}, \citenamefont {Nunnenkamp}, \citenamefont {Walter},\ and\ \citenamefont {Bruder}}]{amitai2017synchronization}%
  \BibitemOpen
  \bibfield  {author} {\bibinfo {author} {\bibfnamefont {E.}~\bibnamefont {Amitai}}, \bibinfo {author} {\bibfnamefont {N.}~\bibnamefont {L{\"o}rch}}, \bibinfo {author} {\bibfnamefont {A.}~\bibnamefont {Nunnenkamp}}, \bibinfo {author} {\bibfnamefont {S.}~\bibnamefont {Walter}},\ and\ \bibinfo {author} {\bibfnamefont {C.}~\bibnamefont {Bruder}},\ }\href@noop {} {\bibfield  {journal} {\bibinfo  {journal} {Physical Review A}\ }\textbf {\bibinfo {volume} {95}},\ \bibinfo {pages} {053858} (\bibinfo {year} {2017})}\BibitemShut {NoStop}%
\bibitem [{\citenamefont {Wang}\ \emph {et~al.}(2014)\citenamefont {Wang}, \citenamefont {Huang}, \citenamefont {Lai},\ and\ \citenamefont {Grebogi}}]{wang2014nonlinear}%
  \BibitemOpen
  \bibfield  {author} {\bibinfo {author} {\bibfnamefont {G.}~\bibnamefont {Wang}}, \bibinfo {author} {\bibfnamefont {L.}~\bibnamefont {Huang}}, \bibinfo {author} {\bibfnamefont {Y.-C.}\ \bibnamefont {Lai}},\ and\ \bibinfo {author} {\bibfnamefont {C.}~\bibnamefont {Grebogi}},\ }\href@noop {} {\bibfield  {journal} {\bibinfo  {journal} {Physical review letters}\ }\textbf {\bibinfo {volume} {112}},\ \bibinfo {pages} {110406} (\bibinfo {year} {2014})}\BibitemShut {NoStop}%
\end{thebibliography}%

\end{document}